\documentclass[aps,twocolumn,showpacs,superscriptaddress,notitlepage,floatfix]{revtex4-2}
\usepackage{amsmath,braket,amssymb,natbib,graphicx,amsthm,xcolor,outlines,mathrsfs,multirow,hhline}
\usepackage[normalem]{ulem} %

\usepackage{dcolumn}%
\usepackage[english]{babel}
\usepackage[utf8x]{inputenc}
\usepackage[T1]{fontenc}
\usepackage{listings}
\usepackage[colorinlistoftodos]{todonotes}
\usepackage[colorlinks=true, allcolors=blue]{hyperref}
\usepackage{enumitem}
\usepackage{booktabs} %
\usepackage{placeins}
\usepackage{hyperref}

\begin{document}
\title{Quantum non-Gaussian coherence via dissipation to single qubits}
\author{Sirachat Kaewkongtham}
\email{sirachat.kaewkongtham@upol.cz}
\affiliation{Department of Optics, Palack\'{y} University, 17. Listopadu 1192/12, 779 00 Olomouc, Czech Republic}
\author{Kimin Park}
\email{park@optics.upol.cz}
\affiliation{Department of Optics, Palack\'{y} University, 17. Listopadu 1192/12, 779 00 Olomouc, Czech Republic}
\author{Darren W. Moore}
\email{darren.moore@upol.cz}
\affiliation{Department of Optics, Palack\'{y} University, 17. Listopadu 1192/12, 779 00 Olomouc, Czech Republic}
\author{Atirach Ritboon}
\email{ritboon@tu.ac.th}
\affiliation{Research Unit in Energy Innovations and Modern Physics (EIMP), Thammasat University, Khlong Nueng, Khlong Nang, Pathum Thani, 12120, Thailand}
\author{Radim Filip}
\email{filip@optics.upol.cz}
\affiliation{Department of Optics, Palack\'{y} University, 17. Listopadu 1192/12, 779 00 Olomouc, Czech Republic}

\date{\today}

\begin{abstract}
  Quantum non-Gaussian (QNG) coherence  is a critical resource for advanced quantum technologies with bosonic systems, operationally distinct from the coherence provided by Gaussian states. For many experimental platforms, its deterministic generation  remains a formidable experimental challenge. 
  Here we propose  that pulsed Gaussian control of a linear oscillator alternating with dissipation by the smallest nonlinear element---an unmeasured ground-state qubit coupled via resonant Jaynes-Cummings (JC) interactions---deterministically generates  high-order QNG coherence.
  Although such dissipative processes are phase-insensitive and do not generate local coherence alone, when added to Gaussian operations they deterministically produce high-quality binary superpositions of lower Fock states \mbox{$(|0\rangle + |N\rangle)/\sqrt{2}$} up to \mbox{$N=4$} with high fidelity (\mbox{$\mathcal{F} > 0.98$}), certified by the QNG coherence hierarchy, as well as binomial code states and logical magic states with high fidelity ($\mathcal{F} > 0.98$).
\end{abstract}

\maketitle

\section{Introduction}

Quantum coherence, the ability of a system to exist in multiple states simultaneously, is a foundational resource in quantum science.
For bosonic systems, coherence in the energy basis is of primary interest \cite{streltsov2017colloquium} and is a prerequisite for many quantum experiments.
Specifically, maximizing off-diagonal Fock-basis coherence surpassing nonclassical \cite{innocentiPRL2023classicalthreshold} or even quantum non-Gaussian (QNG) thresholds ~\cite{Asenbeck2025,lachman_hierarchies_PRA2025,podhora_quantum_2022,kovalenko_quantum_2025} directly and operationally distinguishes genuine high-order non-Gaussian superpositions from Gaussian-transformed lower-order states, as also formalized in the stellar rank hierarchy of non-Gaussian states~\cite{Chabaud2020PRL,Walschaers2021PRXQ}.

For linear oscillators initialized in Gaussian states, deterministically generating quantum non-Gaussian states without measurements strictly requires nonlinear interactions to an auxiliary nonlinear system~\cite{Lloyd1999CV,Weedbrook2012RMP,FilipMista2011PRL,LawEberly1996ArbitraryControl}.
What, then, is the minimal coupling and nonlinearity required to deterministically generate single-mode Fock-basis coherences without measurements or active qubit driving?
      
Surprisingly, towards the preparation of Fock states it was shown that with a smallest nonlinear system---a single saturable qubit---the Jaynes-Cummings (JC) interaction (a purely energy-conserving exchange) alongside thermal oscillator states and ground state qubits is able to produce nonclassical states with large individual Fock components~\cite{slodicka_deterministic_2016,MarekPhysRevA.94.013850nonclassicality} even with enhanced sensitivity \cite{delakouras_production_2024}.
Furthermore, this approach has also been extended to two-mode systems, demonstrating that analogous passive JC mechanisms can induce genuine non-Gaussian entanglement and dual-mode coherence from thermal resources \cite{OELaha2022ThermalEntanglement,LahaPhysRevResearch2024QNGEntanglement}.
Yet, thermal states lack initial phase coherence; hence, passive JC dissipation alone cannot generate single-mode Fock-basis coherences \cite{VogelWelschPhysRevA1989kJC}. We continue this bottom-to-top approach to hierarchically identify the lowest nonlinear resources needed and provide operational benchmarks to justify using higher ones.

In this work, we present a deterministic pulsed protocol that optimally transforms quantum coherence provided by Gaussian states and operations into high-order quantum non-Gaussian coherence in targeted Fock-basis elements. It uses sequential, passive JC interactions that dissipate to a single unmeasured ground-state qubit, alternating with active Gaussian operations. The dissipative mechanism to the single qubit exploits
coherent dynamics and qubit saturation: target Fock components are selectively concentrated 
while intermediate Fock components are lowered via the optimal combination of destructive bosonic interference and dissipation to qubit.  %
It enables the high-fidelity synthesis of single-mode vacuum-Fock superpositions \mbox{$(|0\rangle + |N\rangle)/\sqrt{2}$} up to \mbox{$N=4$}, with a high fidelity susceptibility and importantly, certified by the QNG coherence hierarchy, as well as binomial code states and magic states.

This minimal pulsed dissipation to a single qubit method is applicable over many platforms with the JC model \mbox{\cite{Eberly1980,Rempe1987,Shore1993}}, such as optical cavity QED \mbox{\cite{ReisererRempe2015RevModPhys.87.1379,Welte2017PRL}}, electromechanical resonators \mbox{\cite{chuNature2018multiphononFock,bild_schrodinger_2023}}, and circuit quantum acoustodynamics~\mbox{\cite{Chu2017Science,Satzinger2018Nature,Yang2026Science}}, but also as an alternative to trapped ions \mbox{\cite{podhora_quantum_2022,saner_generating_2024,MatsosPhysRevLett.133.050602}}, superconducting circuits \mbox{\cite{FinkNature2008JC,HofheinzNature2009ArtibrarySynthesis,HollandPhysRevLett2015Fock,LiPhysRevLett2024StabilizationFock}}. %
Using only this dissipation to smallest nonlinear qubit element, we also provide operational benchmarks for other methods that can conclusively provide QNG coherences.

\section{Protocol and Physical Model}

\begin{figure*}[thp]
  \centering
  \includegraphics[width=\textwidth]{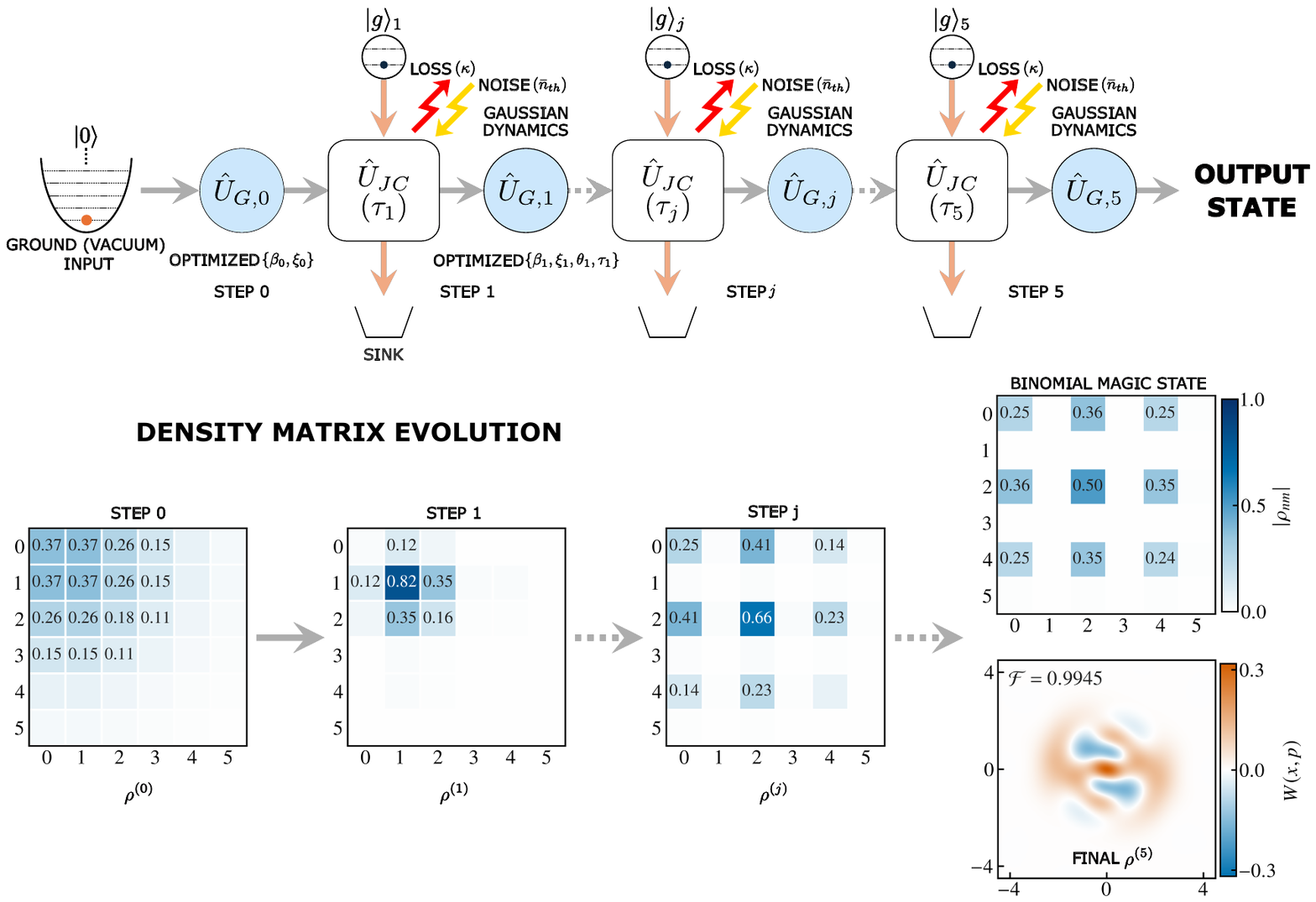}
  \caption{
    \textbf{Deterministic sequential dissipative protocol {using a single qubit} for non-Gaussian coherence preparation.} {(Top) Schematic of the $k$-step cycle:} {after energy-active $U_{G,0}$} an initial Gaussian state {dissipates to} a ground-state qubit ($|g\rangle{_1}$) via {an interaction with} an {energy-passive} JC Hamiltonian for time $t_j$ ($\tau_j = g t_j$). The qubit is discarded and replaced by a new qubit dissipator in ground state, or reset to the ground state,  erasing any qubit-oscillator correlations. %
    A subsequent Gaussian active operation $\hat{U}_{G,j}$ creates destructive modulation in the oscillator to suppress unwanted Fock components{} and reshape the density matrix before the next qubit dissipation (see Table~\ref{tab:parameters} for operational parameters $\{\beta_0, \xi_0, \beta_j, \xi_j, \theta_j, \tau_j\}_{j=1}^k$ preparing $(|0\rangle+|4\rangle)/\sqrt{2}$, while the illustrated bottom panel specifically depicts the step-by-step density matrix evolution toward the binomial magic state $|T\rangle_L$). (Bottom)
    The output bosonic density matrix at each step. The active Gaussian operations and single qubit dissipation deterministically reshape to coherences of a finite superposition of oscillator Fock states (e.g., $|0\rangle+|N\rangle$ and binomial magic states).
  }
  \label{fig:scheme}
\end{figure*}

The protocol prepares target quantum non-Gaussian states via a deterministic sequence of $k=5$ discrete cycles (Fig.~\ref{fig:scheme}). A bosonic mode with annihilation (creation) operator $\hat{a}$ ($\hat{a}^\dagger$), satisfying $[\hat{a},\hat{a}^\dagger]=1$, is initialized in a pure Gaussian state $|\psi_{\text{osc}}(0)\rangle = \hat{S}(\xi_0)\hat{D}(\beta_0)|0\rangle$, where $\hat{D}(\beta) = \exp(\beta\hat{a}^\dagger - \beta^*\hat{a})$ is the displacement operator and $\hat{S}(\xi) = \exp[\frac{1}{2}(\xi^*\hat{a}^2 - \xi\hat{a}^{\dagger 2})]$ is the squeeze operator with complex squeezing parameter $\xi = r e^{i\phi}$. In each cycle $j \in \{1, \dots, k\}$, the oscillator in state $\rho_{\text{osc}}^{(j-1)}$ interacts resonantly with an unmeasured ancilla qubit defined on the two-level basis $\{|g\rangle, |e\rangle\}$ (ground state $|g\rangle$, excited state $|e\rangle$). Initialized in $|g\rangle$, the ancilla couples via the resonant Jaynes-Cummings (JC) Hamiltonian
\begin{equation}
  \hat{H}_{\mathrm{JC}} = g\bigl(\hat{a}\,\hat{\sigma}_{+} + \hat{a}^{\dagger}\hat{\sigma}_{-}\bigr),
  \label{eq:H_JC}
\end{equation}
for duration $t_j$ with coupling strength $g$ and transition operators $\hat{\sigma}_+ = |e\rangle\langle g|$, $\hat{\sigma}_- = |g\rangle\langle e|$. Parameterizing the interaction by the dimensionless duration $\tau_j = gt_j$ makes the unitary evolution and generated states independent of $g$;  duration scales as $t_j = \tau_j/g$. Operating at exact resonance under the rotating-wave approximation, free-energy terms vanish in the interaction picture. We finalize the dissipative process by tracing out the unmeasured qubit (or by active reset), which, followed by an active Gaussian unitary $\hat{U}_{G,j}$, yields the updated oscillator state:
\begin{align}
\rho_{\text{osc}}^{(j)} &= \hat{U}_{G,j}\,\mathrm{Tr}_{\text{qb}}\!\Big[ e^{-i\hat{H}_{\text{JC}} t_j}
  \big( \rho_{\text{osc}}^{(j-1)} \otimes |g\rangle\langle g| \big)
  e^{i\hat{H}_{\text{JC}} t_j} \Big] \hat{U}_{G,j}^{\dagger} \nonumber \\
&= \hat{U}_{G,j} \Big( \sum_{s\in\{g,e\}} \hat{M}_s(\tau_j)\, \rho_{\text{osc}}^{(j-1)}\, \hat{M}_s^{\dagger}(\tau_j) \Big) \hat{U}_{G,j}^{\dagger},
\label{eq:traceout}
\end{align}
with Kraus operators
\begin{align}
  \hat{M}_g(\tau_j) &= \sum_{n=0}^\infty \cos(\tau_j\sqrt{n})\,|n\rangle\langle n|, \label{eq:Mg_main} \\
  \hat{M}_e(\tau_j) &= -i \sum_{n=1}^\infty \sin(\tau_j\sqrt{n})\,|n-1\rangle\langle n|, \label{eq:Me_main}
\end{align}
resolving the joint evolution $e^{-i\hat{H}_{\text{JC}}t}|n, g\rangle = \cos(\tau\sqrt{n})|n,g\rangle - i\sin(\tau\sqrt{n})|n-1,e\rangle$. The Gaussian operations are defined as
\begin{equation}
  \hat{U}_{G,j} = \hat{R}(\theta_j)\hat{S}(\xi_j)\hat{D}(\beta_j),
  \label{eq1:Gaussian}
\end{equation}
with phase rotation $\hat{R}(\theta_j) = \exp(-i\theta_j\hat{n})$, where $\hat{n} = \hat{a}^\dagger\hat{a}$ is the photon-number operator. 
The physical mechanism operates in four stages: (i) Resonant JC exchange entangles the oscillator with the ancilla as $|n, g\rangle \to \cos(\tau_j\sqrt{n})|n,g\rangle - i\sin(\tau_j\sqrt{n})|n-1,e\rangle$. (ii) Intermediate Fock components transfer population into $|e\rangle$ via incommensurate Rabi frequencies $\Omega_n = g\sqrt{n}$. (iii) Tracing out (or resetting) the ancilla irreversibly discards the $|e\rangle$ population via the non-unitary Kraus lowering map $\hat{M}_e$, while $\hat{M}_g$ applies number-dependent amplitude modulation $\cos(\tau_j\sqrt{n})$, corresponding to the transmission operator of a microscopic maser~\cite{Filipowicz1986PRA}. (iv) Interleaved Gaussian operations $\hat{U}_{G,j}$ dynamically re-align amplitudes and phases, driving destructive interference that cancels residual errors across $k=5$ cycles without requiring steady-state dark trapping~\cite{Filipowicz1986PRA} within an actively steered discrete collision model~\cite{Ciccarello2022PhysRep}. Operationally, commuting the phase rotation $\hat{R}(\theta_j) = \exp(-i\theta_j\hat{n})$ forward into subsequent cycles shifts the drive parameters (Appendix~\ref{append:kraus}) via $\hat{D}(\beta_{j'})\hat{R}(\theta_j) = \hat{R}(\theta_j)\hat{D}(\beta_{j'} e^{i\theta_j})$ and $\hat{S}(\xi_{j'})\hat{R}(\theta_j) = \hat{R}(\theta_j)\hat{S}(\xi_{j'} e^{2i\theta_j})$ ($\beta_{j'} \to \beta_{j'} e^{i\theta_j}$, $\xi_{j'} \to \xi_{j'} e^{2i\theta_j}$ for $j' > j$), while the base similarity transformation satisfies $\hat{R}(\theta)\hat{D}(\beta)\hat{R}^\dagger(\theta) = \hat{D}(\beta e^{-i\theta})$ and $\hat{R}(\theta)\hat{S}(\xi)\hat{R}^\dagger(\theta) = \hat{S}(\xi e^{-2i\theta})$; intermediate rotations can thus be absorbed into subsequent drives or deferred to the final step $j = k$, simplifying intermediate control requirements without compromising state fidelity. The complex parameters of displacement $\beta_j$ and squeezing $\xi_j$ enable this absorption directly. Because phase rotations commute with the Jaynes--Cummings Kraus channel ($[\hat{R}_\theta, \mathcal{E}_{\text{JC}}] = 0$) and leave the vacuum invariant ($\hat{R}_\theta|0\rangle = |0\rangle$), commuting rotations backward through the sequence absorbs them into earlier drive phases. Physical phase-rotation gates are therefore redundant in hardware; explicit $\hat{R}(\theta_j)$ is retained during optimization solely to avoid coordinate singularities when squeezing vanishes ($r_j \to 0$, Appendix~\ref{append:kraus}). For the inter-step dynamics, we assume either a fresh ground-state qubit each cycle or a fast active reset mechanism (\mbox{$t_{\text{reset}} \ll t_j$}), preventing oscillator decoherence during idle periods, and neglect intermediate losses during Gaussian operations under the unitary approximation in Eq.~(\ref{eq1:Gaussian}). We consider linear Gaussian operations to be ideal and focus on the impact of the nonlinear dissipative interaction.
All operational parameters $\{\beta_0, \xi_0, \beta_j, \xi_j, \theta_j, \tau_j\}_{j=1}^k$ are globally optimized across all $k$ steps simultaneously to maximize the fidelity $\mathcal{F} = \langle \psi_{\text{target}} | \rho_{\text{osc}}^{(k)} | \psi_{\text{target}} \rangle$ with the target state $|\psi_{\text{target}}\rangle$
 (Details are in Appendix~\ref{append:params}, Table~\ref{tab:parameters}, and coherence flow in Appendix~\ref{append:coherence_flow}, Table~\ref{tab:coherence_values}).

\subsection{Quantum Non-Gaussian Coherence}

To specifically verify the genuine quantum non-Gaussian coherence of the specific target QNG coherence beyond Wigner negativity, we analyze the evolution of genuine quantum non-Gaussian coherence (QNG) using the Fock-space hierarchy criteria \cite{LachmanPRL2019Hierarchy,kovalenko_quantum_2025,Asenbeck2025}.
We quantify the coherence between Fock-state components $n$ and $m$ directly via the off-diagonal elements of the density matrix $\rho$ in the Fock basis, defined as $C_{n,m} = 2|\rho_{nm}|$. This directly benchmarks our binary superpositions $(|0\rangle + |N\rangle)/\sqrt{2}$ and $(|1\rangle + |N\rangle)/\sqrt{2}$; multi-component states (such as the binomial magic state illustrated in Fig.~\ref{fig:scheme}) are evaluated via state fidelity in Sec.~\ref{sec:binomial}.
Here, $C_G \equiv \sup_{\rho \in \mathcal{G}} 2|\langle n|\rho|m\rangle|$ defines the minimal quantum non-Gaussian threshold as the supremum over all convex mixtures of Gaussian states $\mathcal{G}$~\cite{Asenbeck2025}.
The fundamental thresholds correspond to:
\begin{enumerate}
  \item \textbf{Non-classicality:} The bound separating values of $C_{n,m}$ not reachable by classical mixtures of coherent states.
  \item \textbf{Minimal QNG:} The threshold required to prove that a value of $C_{n,m}$ cannot be represented as a mixture of Gaussian states.
  \item \textbf{QNG Hierarchy:} Stronger criteria which verify that $C_{n,m}$ cannot be reached using Gaussian operations on lower-order QNG coherences, in operational correspondence with the stellar hierarchy of non-Gaussian states where $(|0\rangle+|N\rangle)/\sqrt{2}$ has stellar rank $r=N$~\cite{Chabaud2020PRL,Walschaers2021PRXQ}.
\end{enumerate}

Under single-photon loss $\kappa$ and thermal noise $\bar{n}_{\text{th}}$, we quantify robustness by tracking the decay of coherence $|\rho_{nm}(t)|$. The critical times $t_{\text{crit}}^{\text{gen}}$ and $t_{\text{crit}}^{\text{ideal}}$ denote when the coherence of the generated and ideal target states cross below a given QNG threshold. The ratio $t_{\text{crit}}^{\text{gen}}/t_{\text{crit}}^{\text{ideal}}$ quantifies the resilience of the deterministically prepared state relative to the ideal theoretical benchmark (decay dynamics detailed in Appendix~\ref{appendix:decay}).

\section{Results}

We first evaluate the protocol under ideal, noise-free conditions before analyzing robustness against experimental decoherence. The complete set of optimized parameters is provided in Appendix~\ref{append:params}.

\subsection{Binary Fock Superpositions \& QNG Coherence Hierarchy}

\begin{figure*}[t]
  \centering
  \includegraphics[width=\textwidth]{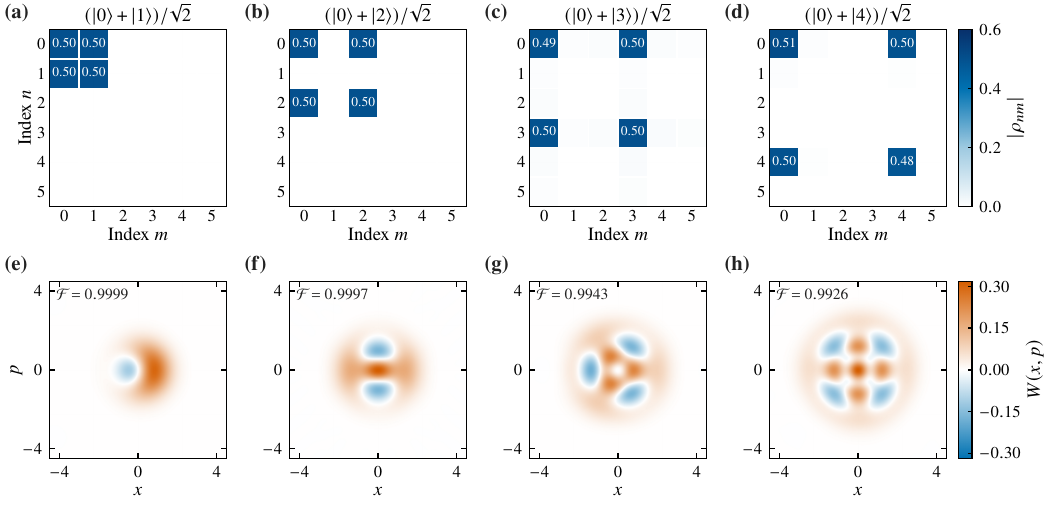}
  \caption{\textbf{Deterministic dissipative preparation of $(|0\rangle + |N\rangle)/\sqrt{2}$ Fock state superpositions via $k=5$ sequential Jaynes--Cummings (JC) interactions.} (a--h) State characterization for $N=1, 2, 3,$ and $4$ (left to right). Panels (a--d) display the absolute values of the density matrix elements $|\rho_{nm}|$ in the Fock basis for the generated states; the states are normalized over the full Fock space, and only the block $0 \le n, m \le 5$ is shown. Panels (e--h) show the corresponding Wigner functions $W(x, p)$, exhibiting Wigner negativity characteristic of the target quantum non-Gaussian states in phase space. Each state is shown after the free phase rotation $e^{i\theta\hat{n}}$ that makes $\rho_{0N}$ real and positive; this fixes the orientation of $W$ without changing $|\rho_{nm}|$, $\mathcal{F}$ or the purity. 
  The protocol achieves high fidelities $\mathcal{F}$ relative to the ideal target states $(|0\rangle + |N\rangle)/\sqrt{2}$, specifically $\mathcal{F} = 0.9999$ (purity, defined as $\mathrm{Tr}[\rho^2]$, $0.9999$) for $N = 1$, $\mathcal{F} = 0.9997$ (purity $0.9994$) for $N = 2$, $\mathcal{F} = 0.9943$ (purity $0.9911$) for $N = 3$, and $\mathcal{F} = 0.9926$ (purity $0.9872$) for $N = 4$.
  }
  \label{fig:0N}
\end{figure*}

We demonstrate the deterministic synthesis of high-fidelity vacuum-Fock superpositions $(|0\rangle + |N\rangle)/\sqrt{2}$ ($N=1\text{--}4$) in Fig.~\ref{fig:0N}.
The density matrices in Figs.~\ref{fig:0N}(a--d) exhibit a binary structure of the target Fock components $|0\rangle$ and $|N\rangle$  populated with near-equality, while intermediate Fock states are effectively suppressed, disproving a simple diffusive redistribution of Gaussian energy.
High off-diagonal coherence values $|\rho_{0N}|$ confirm successful QNG coherence generation.

The lossless and noiseless protocol achieves high fidelities $\mathcal{F}$ with respect to the ideal target states---specifically $\mathcal{F} = 0.9999, 0.9997, 0.9943$, and $0.9926$ for $N=1$ to $4$, respectively (Fig.~\ref{fig:0N})---demonstrating that $k=5$ interaction steps reach a step-efficient plateau sufficient to deterministically prepare these binary superpositions with high fidelity.
The Wigner functions in Figs.~\ref{fig:0N}(e--h) exhibit pronounced negativity and high-visibility modulation fringes that reproduce those of the target states{~\cite{DelegliseNature2008}}.

To further extend the fidelity evaluation of the generated $(|0\rangle + |N\rangle)/\sqrt{2}$ superpositions and involve their fidelity susceptibility as another coherence quantifier \cite{LuisOe2012FisherInformationCoherence,Yu2022PhysRevA.106.052432Coherence,Li2021PhysRevA.103.012401CoherenceFisher,Tan2018PhysRevA.97.052304CoherenceQFI,Feng2017SciRepQuantifying,tan2021PRLfisher}, we analyzed the normalized infidelity $1-\mathcal{F}(\rho_{\theta},\rho_{\theta+\delta\theta})$ between the exact state $\rho_{\theta}$ and the infinitesimally phase shifted state $\rho_{\theta+\delta\theta} \equiv e^{-i\delta\theta\hat{n}}\rho_{\theta}e^{i\delta\theta\hat{n}}$, which corresponds to the quantum Fisher information (QFI) via the curvature $F_Q = -2 \left.\frac{\partial^2 \mathcal{F}}{\partial (\delta\theta)^2}\right|_{\delta\theta=0}$~\mbox{\cite{BraunsteinCaves1994PRL,Paris2009IJQI,You2007PRE,Zanardi2007PRL,Gu2010IJMPB}}. This quantifies the quantum coherence in terms of the state's intrinsic sensitivity of the off-diagonal elements in the Fock basis to phase shifts generated by the photon number operator $\hat{n}$~\cite{Rahman2025PhysRevLett.134.180801}. The phase-dependent sensitivity for the full hierarchy $N=1, \dots, 4$ is illustrated in Appendix~\ref{append:Fisher} (Supplementary Fig.~\ref{fig:fisher}), where the classical Fisher information (CFI)  defined in Appendix~\ref{append:Fisher} for ideal state projective detection and the corresponding QFI of the generated mixed states are shown.

The CFI oscillates with exactly $N$ characteristic dips to zero. For our fixed optimal projection $\hat{\Pi}_{0N} = |\psi_{0N}\rangle\langle\psi_{0N}|$ onto the ideal state $|\psi_{0N}\rangle = (|0\rangle+|N\rangle)/\sqrt{2}$ \mbox{\cite{vlastakis_deterministically_2013,Heeres2015,HuNatPhys2019ExperimentalBinomial}}, the measurement probability $p(\varphi) = \mathrm{Tr}[\hat{\Pi}_{0N}\rho(\varphi)]$ forms modulation fringes $p(\varphi) \approx \frac{1}{2}(1 + V \cos(N\varphi))$, where $V < 1$ is the visibility limited by the state's slight mixedness and residual Fock populations not visible in the target off-diagonal elements $|\langle 0|\rho|N\rangle|$. The CFI plunges to zero at exactly $\varphi = m\pi/N$ (with integer $m$), while non-unit purity smooths these singularities, causing the CFI to form broad plateaus that approach the QFI maximum between the dips.

For the ideal superpositions $(|0\rangle+|N\rangle)/\sqrt{2}$, the number variance \mbox{$(\Delta n)^2 = N^2/4$} sets the phase-sensitivity bound \mbox{$F_Q^{\mathrm{ideal}} = N^2$}~\mbox{\cite{MonrasPRA2006,Rahman2025PhysRevLett.134.180801,Giovannetti2004Science}}, yielding ideal bounds of $1.00$, $4.00$, $9.00$, and $16.00$ for \mbox{$N=1,2,3,4$}. The generated states reach CFI values of \mbox{$F_C^{\max} = 1.00$} ($N=1$), $4.00$ ($N=2$), $8.88$ ($N=3$), and $15.81$ ($N=4$), yielding extraction ratios of \mbox{$F_C^{\max}/F_Q^{\mathrm{ideal}} = 0.9999$} ($N=1$), $0.9994$ ($N=2$), $0.9869$ ($N=3$), and $0.9882$ ($N=4$). The complete CFI profiles and plateau are in Supplementary Fig.~\ref{fig:fisher}. These high ratios confirm that the generated states efficiently concentrate their non-Gaussian coherence into  phase sensitivity, capturing over $98.6\%$ of the maximum across all orders.

\begin{figure}[t]
\centering
\includegraphics[width=\columnwidth]{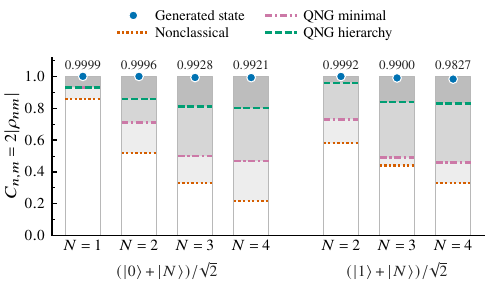}
\caption{\textbf{Coherence values of the generated states $(|0\rangle + |N\rangle)/\sqrt{2}$ and $(|1\rangle + |N\rangle)/\sqrt{2}$ compared to thresholds:} $C_{0,N}=2|\rho_{0N}|$ for the generated states $(|0\rangle + |N\rangle)/\sqrt{2}$ ($N=1$--$4$, left) and $C_{1,N}=2|\rho_{1N}|$ for the generated states $(|1\rangle + |N\rangle)/\sqrt{2}$ ($N=2$--$4$, right), obtained after five Jaynes--Cummings interactions. Within each column, the horizontal marks represent the limits for nonclassical coherence (dotted), the minimal quantum non-Gaussian threshold (dash-dotted), and the hierarchy of genuine quantum non-Gaussian coherence thresholds (dashed), and the shading darkens above each threshold; for $N=1$ the two QNG thresholds coincide. Dots mark the generated states, with their values above ($C=1$ for the ideal states). All generated states surpass the QNG thresholds, confirming the successful generation of genuine quantum resources. Under photon loss, the generated nonclassicality survives up to loss fractions $L_{\rm crit} = 26.0\%$ ($N=1$) and $53.1\%$ ($N=4$), and all states retain $\ge 90\%$ (with $(|0\rangle+|N\rangle)/\sqrt{2}$ retaining $\ge 95\%$) of the ideal coherence lifetime under thermal decoherence ($\kappa=0.1, \bar{n}_{\rm th}=0.1$; see Appendix~\ref{appendix:decay}, Fig.~\ref{fig:decoherence0N} and Fig.~\ref{fig:thermal_ratio}).}
\label{fig:coherence_thresholds}
\end{figure}
Figure~\ref{fig:coherence_thresholds} summarizes the initial coherence values ($C_{0,N}$ and $C_{1,N}$) of the generated states and compares them directly against the non-classical, minimal QNG, and QNG hierarchy thresholds. All generated states surpass the higher-order QNG hierarchy thresholds, demonstrating that the minimal sequential protocol successfully establishes genuine quantum non-Gaussian coherence up to $N=4$.
Additionally, we investigate asymmetric variants of non-Gaussian states \mbox{$\frac{1}{\sqrt{1+\epsilon^2}}\left(|0\rangle + \epsilon|N\rangle\right)$}. Their non-Gaussian coherence can be tailored between unbalanced superposed states. For this class of states, the QFI \mbox{$F_Q = 4\bar{n}(N-\bar{n})$} with \mbox{$\bar{n} = \epsilon^2 N/(1+\epsilon^2)$} strictly surpasses the Gaussian phase sensitivity bound \mbox{$F_Q^G(\bar{n}) = 8\bar{n}(\bar{n}+1)$}~\cite{MonrasPRA2006} when \mbox{$\epsilon < \sqrt{\frac{N-2}{2N+2}}$}, a condition physically valid for all \mbox{$N \ge 3$}. 
For \mbox{$N=4$}, this threshold is \mbox{$\epsilon < 1/\sqrt{5} \approx 0.447$}; the generated state achieves \mbox{$F_Q/\bar{n} = 15.7$} at $\epsilon = 0.05$, strictly exceeding the Gaussian bound \mbox{$F_Q^G/\bar{n} = 8(\bar{n}+1) \approx 8.08$} while strictly conforming to the 4-photon subspace limit (\mbox{$F_Q/\bar{n} = 4(\Delta n)^2/\bar{n} \le 4N = 16$} for any state supported on $n \le 4$). State sensitivity is therefore certified above the minimal QNG threshold.

The phase sensitivity of the generated states is encoded in the corresponding off-diagonal elements of the Fock basis. These quantum coherences can be individually certified as achievable only by quantum non-Gaussianity~\mbox{\cite{Asenbeck2025,kovalenko_quantum_2025}}. The generated states simultaneously achieve high genuine non-Gaussian coherence: \mbox{$2|\rho_{01}| = 0.9999$}, \mbox{$2|\rho_{02}| = 0.9996$}, \mbox{$2|\rho_{03}| = 0.9928$}, and \mbox{$2|\rho_{04}| = 0.9921$} (Appendix~\ref{append:params}, Fig.~\ref{fig:Evol}), all decisively surpassing the minimal Gaussian mixture thresholds (\mbox{$C_G \approx 0.47$} for minimal QNG and \mbox{$C_{\text{hier}} = 0.80$} for the highest QNG hierarchy at \mbox{$N=4$}). Under photon loss, the off-diagonal coherence $|\rho_{0N}|$ remains above the nonclassicality threshold up to critical loss fractions of $26.0\%$ ($N=1$), $48.0\%$ ($N=2$), $52.4\%$ ($N=3$), and $53.1\%$ ($N=4$) [Supplementary Fig.~\ref{fig:decoherence0N}], within $0.5$ percentage points of the ideal states. Furthermore, under single-photon loss (\mbox{$\kappa$}) and thermal noise (\mbox{$\bar{n}_{\text{th}} = 0.1$}), the ratio of critical survival times \mbox{$t_{\text{crit}}^{\text{gen}}/t_{\text{crit}}^{\text{ideal}}$} retains over \mbox{$90\%$} of the ideal state's persistence at the highest QNG hierarchy threshold (Supplementary Fig.~\ref{fig:thermal_ratio} and Appendix~\ref{appendix:decay}), indicating that the decay is dominated by the intrinsic fragility of the target superposition rather than by residual preparation imperfections.
To demonstrate that other QNG coherences are achievable, Appendix~\ref{appendix:appendix1N} details the synthesis of non-vacuum binary superpositions $(|1\rangle + |N\rangle)/\sqrt{2}$ ($N=2,3,4$), confirming that the protocol maintains high-fidelity coherence and QNG thresholds consistent with the vacuum-superposition results.

\subsection{Binomial Codes and Logical Magic States}
\label{sec:binomial}

Beyond binary superpositions, in the lowest-order binomial code \cite{MichaelPRX2016Binomial,Albert2018PRA}, quantum information is protected against single-photon loss by distributing amplitude across even-parity Fock states $\{\ket{0},\ket{2},\ket{4}\}$. While an even binary superposition like $(|0\rangle+|4\rangle)/\sqrt{2}$ collapses under single-photon loss into a single Fock state $\hat{a}(|0\rangle+|4\rangle)/\sqrt{2} = \sqrt{2}|3\rangle$ (completely destroying coherence), the target symmetric even code state $|+_L\rangle \equiv (|0_L\rangle + |1_L\rangle)/\sqrt{2} = \frac{1}{2}(|0\rangle+|4\rangle) + \frac{1}{\sqrt{2}}|2\rangle$ (with logical basis $|0_L\rangle \equiv (|0\rangle+|4\rangle)/\sqrt{2}$ and $|1_L\rangle \equiv |2\rangle$) transforms under loss to $\hat{a}|+_L\rangle = |1\rangle + |3\rangle$. This single-photon loss state resides strictly within the odd-parity error subspace $\mathcal{H}_{\mathrm{err}} = \mathrm{span}\{|1\rangle, |3\rangle\}$, orthogonal to the even logical subspace. Measuring photon-number parity $\hat{\Pi} \equiv (-1)^{\hat{n}}$ unambiguously detects the error without destroying the coherent superposition between the error basis states, enabling deterministic recovery~\mbox{\cite{Sun2014Tracking,HuNatPhys2019ExperimentalBinomial}}.

 Preparing this multi-component superposition dissipatively presents a fundamental obstacle: because the Jaynes--Cummings Rabi frequencies $\Omega_n = g\sqrt{n}$ have mutually irrational ratios (e.g., $\Omega_2/\Omega_1 = \sqrt{2}$ and $\Omega_4/\Omega_2 = \sqrt{2}$), no single unsteered interaction duration $\tau_j = gt_j$ can simultaneously preserve desired even Fock components ($\sin(\tau_j\sqrt{2}) \approx 0$, $\sin(2\tau_j) \approx 0$) while extinguishing intermediate odd populations ($n=1, 3$). Unlike continuous dissipative engineering constrained by asymptotic decay and dark-state trapping, our pulsed protocol iteratively clears residual errors via destructive interference across \mbox{$k=5$} steps (Appendix~\ref{append:kraus}), establishing the necessity of multiple steps.

Beyond standard logical encoding, we apply the protocol to generate target maximal-coherence logical magic states \mbox{$|T\rangle_L \equiv (\vert{}0\rangle_L + e^{i\pi/4}\vert{}1\rangle_L)/\sqrt{2}$} and \mbox{$|H\rangle_L \equiv \cos(\pi/8)|0_L\rangle + \sin(\pi/8)|1_L\rangle$} in the binomial code basis~\cite{MichaelPRX2016Binomial,HuNatPhys2019ExperimentalBinomial}, defined directly as \mbox{$|0_L\rangle = \frac{1}{\sqrt{2}}(|0\rangle+|4\rangle)$} and \mbox{$|1_L\rangle = |2\rangle$} across the even-parity support \mbox{$\{|0\rangle, |2\rangle, |4\rangle\}$}. These resources are required as ancillas to implement non-Clifford gates for universal fault-tolerant quantum computation with rotation-symmetric bosonic codes~\cite{Grimsmo2020RSBC}, as standard error-correcting codes cannot transversally implement universal gate sets. 

\begin{figure*}
  \centering
  \includegraphics[width=0.92\textwidth]{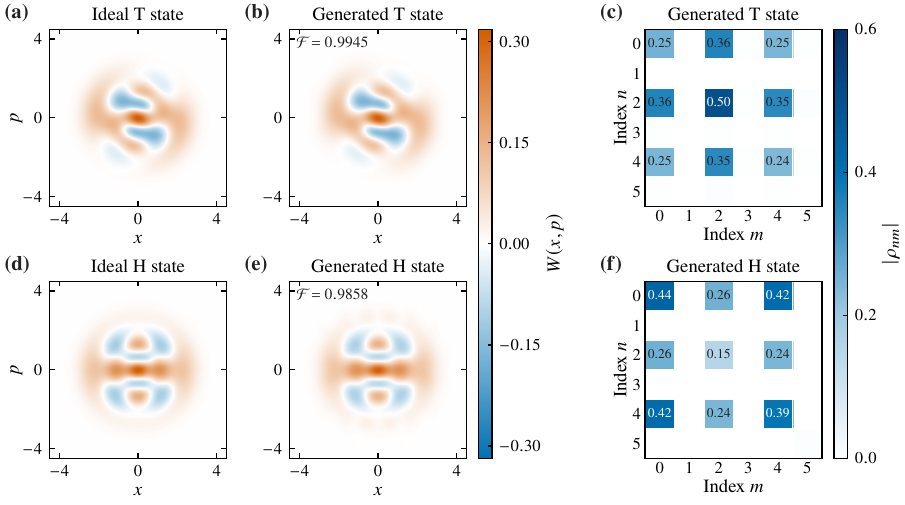}

   \caption{\textbf{Generation of logical magic states in the binomial code basis.} (a) and (d) Wigner functions of the ideal logical $T$-state and $H$-state, respectively; (b) and (e) Wigner functions of the generated $T$-state and $H$-state, respectively, after the optimized protocol. (c) and (f) Absolute values of the density matrix elements $|\rho_{nm}|$ in the Fock basis for the generated $T$-state and $H$-state, respectively. The generated $T$-state achieves a state overlap fidelity of $\mathcal{F} = 0.9945$ (purity $0.9903$), while the generated $H$-state reaches $\mathcal{F} = 0.9858$ (purity $0.9753$). The fidelities include the free phase rotation $e^{i\theta\hat{n}}$, which also sets the orientation of the generated Wigner functions. The states are computed with a Fock cutoff of $40$, and the density matrices are shown for $0 \le n, m \le 5$ to encompass the dominant populations and residual odd-parity leakage of the generated states beyond the even code manifold $\{|0\rangle, |2\rangle, |4\rangle\}$.}
  \label{fig:Magicstates}
\end{figure*}

While pairwise QNG coherence certifies individual off-diagonal superpositions, multi-component code and magic states require global density matrix certification. We therefore benchmark these states using state fidelity.
The synthesized  states achieve state overlap fidelities \mbox{$\mathcal{F} \equiv \langle \psi|\rho|\psi\rangle$} of \mbox{$\mathcal{F} = 0.9945$} (purity $0.9903$) for \mbox{$|T\rangle_L$} and \mbox{$\mathcal{F} = 0.9858$} (purity $0.9753$) for \mbox{$|H\rangle_L$} (Fig.~\ref{fig:Magicstates}).  Because the target states reside entirely within the logical code space with projector $\hat{P}_L \equiv |0_L\rangle\langle 0_L| + |1_L\rangle\langle 1_L|$ ($\hat{P}_L |\psi\rangle = |\psi\rangle$), projecting onto the code space yields a decoded logical fidelity $\mathcal{F}_{\text{logical}} = \mathcal{F} / P_{\text{code}} \ge \mathcal{F}$ (with code occupancy $P_{\text{code}} = \mathrm{Tr}[\hat{P}_L \rho] > 98.5\%$). 

  In fault-tolerant architectures with rotation-symmetric bosonic codes, 
  odd-parity leakage is heralded by parity measurement $\hat{\Pi}$~\cite{Grimsmo2020RSBC,ChamberlandPRXQ2022}, where active coherent ancilla control enables non-classical state synthesis and universal gate operations, including non-Clifford gates in binomial logical qubits~\cite{HuNatPhys2019ExperimentalBinomial}.
To theoretically compare, we use the Bravyi--Kitaev distillation fidelity threshold $\mathcal{F}\approx 0.859$~\cite{BravyiPRA2005} for both states $|T\rangle_L$ and $|H\rangle_L$. Both the state overlap fidelities ($\mathcal{F} > 0.985$) and decoded logical fidelities ($\mathcal{F}_{\text{logical}} > 0.99$) exceed this threshold. Because their infidelities ($1 - \mathcal{F} < 0.015$, with decoded logical infidelity $1 - \mathcal{F}_{\text{logical}} < 0.01$) are an order of magnitude below the distillation error threshold ($1 - \mathcal{F}_{\text{th}} \approx 0.141$), these prepared states qualify directly as high-fidelity input resources for fault-tolerant magic-state distillation without requiring initial purification stages. While magic state distillation protocols have been established theoretically for bosonic codes (notably GKP codes~\cite{Baragiola2019PRL,David2026arXiv}), experimental demonstrations of distillation have so far been performed only in discrete physical or logical qubits~\cite{Souza2011,Ye2023,Kim2024}. %
Our open-system simulations confirm that the synthesized states remain  $\mathcal{F} > 0.859$ up to $7.3\%$ photon loss and retain $>90\%$ of the ideal thermal lifetime (Supplementary Fig.~\ref{fig:thermal_ratio} and Fig.~\ref{fig:decoherence_TH}).
Crucially, continuous-variable bosonic states mapped to encoded qubits satisfy rigorous sufficient conditions for universal quantum computation via magic state distillation~\cite{CalcluthPRXQ2024}.

\begin{figure*}
  \centering
  \includegraphics[width=\textwidth]{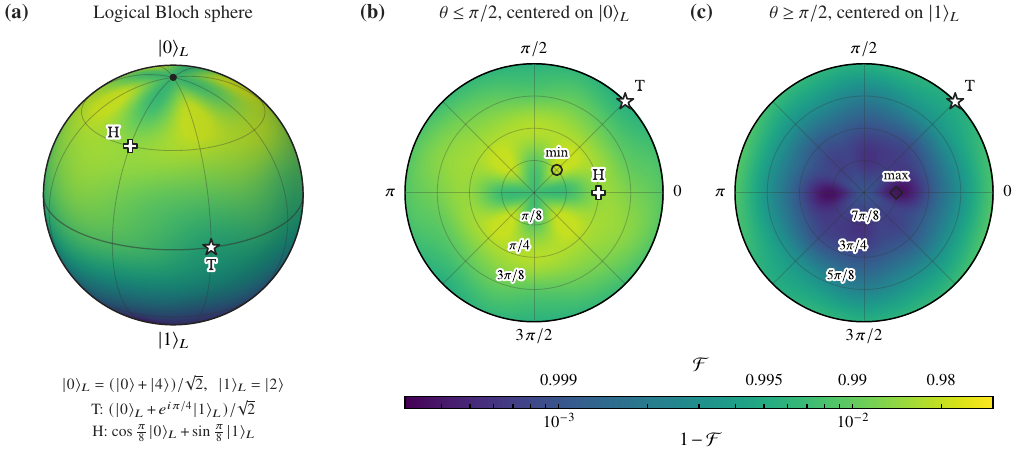}
    \caption{\textbf{Protocol performance across the logical Bloch sphere of the binomial code.} Infidelity $1-\mathcal{F}$ of the optimized $k=5$ protocol for every target state $|\psi(\theta,\phi)\rangle=\cos(\theta/2)|0\rangle_L+e^{i\phi}\sin(\theta/2)|1\rangle_L$, with $|0\rangle_L=(|0\rangle+|4\rangle)/\sqrt{2}$ and $|1\rangle_L=|2\rangle$, shown on a logarithmic color scale (top axis: $\mathcal{F}$). (a) Logical Bloch sphere, with the logical $T$- and $H$-states marked by a star and a cross. (b), (c) Northern ($\theta\le\pi/2$) and southern ($\theta\ge\pi/2$) hemispheres projected onto discs centered on $|0\rangle_L$ and $|1\rangle_L$, respectively, both viewed from above ($+z$);  radial coordinate is polar distance from the respective pole and the angle is $\phi$. The protocol was optimized independently at 58 points of a grid with $\theta=m\pi/8$ and $\phi=p\pi/8$; interpolated bicubically. At the three grid points that coincide with individually optimized states ($|0\rangle_L$, $T$ and $H$), the solutions of those optimizations are used. %
  Half-turn symmetry $\mathcal{F}(\theta,\phi)=\mathcal{F}(\theta,\phi+\pi)$ arises because the phase-space rotation $e^{i\pi\hat{n}/2}$  acts as a logical $Z$. %
  The fidelity exceeds $0.97$ globally, from $\mathcal{F}=0.9799$ at $(\theta,\phi)=(\pi/8,\pi/4)$ (circle) to $\mathcal{F}=0.9996$ at $(7\pi/8,0)$ (diamond); the hardest targets lie on the $\theta=\pi/8$ and $\pi/4$ rings close to $|0\rangle_L$, while states dominated by $|1\rangle_L=|2\rangle$ are generated most accurately. The $T$- and $H$-states reach $\mathcal{F}=0.9945$ and $0.9858$, and the pole $|0\rangle_L$ reaches $\mathcal{F}=0.9926$, identical to the individually optimized states (Fig.~\ref{fig:0N} for $|0\rangle_L$ and Fig.~\ref{fig:Magicstates} for $T$ and $H$). All fidelities include the free final phase-space rotation and are computed in a Fock space truncated at $n_\mathrm{max}=40$, as for the individual states; doubling it to $n_\mathrm{max}=80$ changes them by at most $10^{-6}$.}
  \label{fig:BlochSphere}
\end{figure*}

The resulting binomial magic states are visualized in Fig.~\ref{fig:Magicstates}. The generated Wigner functions (Fig.~\ref{fig:Magicstates}b,e) reproduce the ideal patterns (Fig.~\ref{fig:Magicstates}a,d), while the density matrices (Fig.~\ref{fig:Magicstates}c,f) confirm the exact even parity structures required for fault-tolerant magic-state preparation.
Beyond discrete points, the protocol maintains high performance across the entire logical manifold: the fidelity exceeds $\mathcal{F} > 0.97$ over the complete logical Bloch sphere of the binomial code (Fig.~\ref{fig:BlochSphere}).

\section{Conclusion, Discussion and Outlook}
\label{sec:discussion}

Our pulsed protocol upgrades Gaussian states and operations to deterministically create high QNG coherences by a dissipation to an unmeasured ground-state qubit via an energy-conserving Jaynes-Cummings interaction, without external coherent qubit preparation, classical drive or measurement.
The coherent dynamics and saturability of the qubit provide the essential nonlinearity via \mbox{$\sqrt{n}$}-dependent Rabi oscillations~\cite{FinkNature2008JC}, enabling selective dissipative modulation of oscillator coherences to deterministically generate high-order quantum non-Gaussian superpositions.
Such QNG coherences are established resources for quantum sensing \mbox{\cite{mccormick_quantum-enhanced_Nature2019,Wolf2019MotionalFock}}, quantum memories \mbox{\cite{Bozkurt2025MechanicalQuantumMemory}}, and bosonic quantum error correction (QEC) \mbox{\cite{MichaelPRX2016Binomial,HuNatPhys2019ExperimentalBinomial,MatsosPhysRevLett.133.050602}}, as well as universal non-Gaussian processing~\cite{Grimsmo2020RSBC}, across superconducting circuits \mbox{\cite{HofheinzNature2009ArtibrarySynthesis,Heeres2015}}, trapped ions \mbox{\cite{saner_generating_2024,kovalenko_quantum_2025}}, cavity QED \mbox{\cite{ReisererRempe2015RevModPhys.87.1379}}\mbox{\cite{Welte2017PRL}}, and atomic ensembles \mbox{\cite{LuckeScience2011TwinFock,pezze2019PRLheralded}}.

Regarding interaction parameters, the global optimum is obtained from searches with wide interaction bounds ($\tau_j \le 80$; for typical circuit QED couplings $g/2\pi = 20\text{ MHz}$, $\tau_j \le 80$ corresponds to physical durations $t_j \le 637\text{ ns}$, well within cavity coherence lifetimes $T_1 \sim 1\text{ ms}$) reaching $\mathcal{F} = 0.9926$ (Table~\ref{tab:parameters}), which exploits quasi-revivals ($\cos(gt_j\sqrt{n}) \approx \pm 1$). While the present parameters employ extended interaction times ($\tau_1 \approx 59.9$, $\tau_3 \approx 68.9$), restricting the pulse duration to the short-interaction regime (e.g., enforcing $\tau_j \le 5.0$, corresponding to $t_j < 40\text{ ns}$ at $g/2\pi = 20\text{ MHz}$ to operate well within realistic ancilla inhomogeneous dephasing limits $T_2^*$~\mbox{\cite{Krantz2019,Blais2021RMP,Place2021,Paik2011PRL}}), %
opens possible studies of experimental feasibility across physical architectures.

 We emphasize that although the transient Jaynes--Cummings coupling entangles the oscillator with the ancilla during each interaction, at the optimized times $\tau_j$ the two nearly factorize, so each ancilla reset costs little purity: for the $(|0\rangle+|4\rangle)/\sqrt{2}$ state, $\mathrm{Tr}[\rho_{\mathrm{osc}}^2]$ stays above $0.996$ for the first four cycles and ends at $0.987$ after the final step. As detailed in Appendix~\ref{append:coherence_flow}, this occurs because the non-unitary ancilla reset breaks joint correlations without information backflow, and the conditional states associated with the Kraus operators asymptotically factorize due to near-collinearity at step $k=5$.

 It is instructive to contrast this protocol with standard continuous dissipative state engineering (DSE)~\mbox{\cite{Poyatos1996PRL,Verstraete2009,de_moraes_neto_dissipative_2019,brunelli_dissipative_2019,rivera_creating_2023}}. In continuous DSE, target states are stabilized as autonomous steady states by coupling the oscillator to engineered reservoirs via continuous drives, a paradigm successfully realized to autonomously stabilize single- and multi-photon Fock states in superconducting microwave cavities~\mbox{\cite{HollandPhysRevLett2015Fock,LiPhysRevLett2024StabilizationFock}} and motional Fock states in trapped ions~\mbox{\cite{KienzlerScience2015Reservoir,kouzelis_dissipative_2020}}. Autonomous continuous stabilization of superposition states and code manifolds in the Fock basis generally relies on engineered reservoirs with multi-photon jump operators~\mbox{\cite{Mirrahimi2014NJP,Leghtas2015Science,kouzelis_dissipative_2020,zeng2023approximate,VanselowPRX2026DissipatingQuartets}}. In contrast, our approach operates out of equilibrium via a pulsed sequence of $k$ discrete cycles~\mbox{\cite{Puente2024}}, combining the simplest passive, energy-conserving Jaynes-Cummings exchange with an unmeasured ground-state qubit and active Gaussian operations. By harnessing destructive interference across successive steps, high-fidelity Fock superpositions and binomial codes are synthesized within finite interaction times without requiring continuous multi-photon drives or engineered reservoir couplings. Furthermore, while conventional Law-Eberly state synthesis~\cite{LawEberly1996ArbitraryControl}, as well as dispersive schemes like SNAP gates~\cite{Krastanov2015PRA} or echoed conditional displacements (ECD)~\cite{Eickbusch2022NatPhys}, relies on alternating coherent qubit driving and qubit-oscillator entangling operations, our scheme replaces active qubit operations with passive dissipation to ground-state qubits, using Gaussian oscillator operations to achieve state synthesis without coherent ancilla control. By adjusting the parameter set \mbox{$\{\beta_0, \xi_0, \beta_j, \xi_j, \theta_j, \tau_j\}_{j=1}^k$}, the system switches between symmetric binomial codes and asymmetric magic states without hardware reconfiguration.

These results establish an alternative baseline across multiple physical architectures where the Jaynes-Cummings interaction is naturally realized.
If the qubit itself were coherently driven or prepared in a coherent superposition (which would already constitute a source of quantum non-Gaussian coherence), a passive JC swap would trivially transfer it to the oscillator. In contrast, our protocol operates with the qubit initialized strictly in the ground state without coherent qubit preparation or classical drives, proving that QNG coherence can be generated solely from the oscillator's Gaussian coherence using passive interactions and resets.
In optical cavity QED, atom-by-atom exchange naturally provides disposable ground-state ancillas~\cite{ReisererRempe2015RevModPhys.87.1379, Welte2017PRL}, while Rydberg superatoms in fiber cavities offer strong two-level saturation~\cite{Vaneecloo2022}. In quantum acoustics (such as high-overtone bulk acoustic wave resonators, HBAR)~\cite{chuNature2018multiphononFock,bild_schrodinger_2023} and trapped-ion motional modes~\cite{podhora_quantum_2022,saner_generating_2024}, long bosonic coherence times combined with resonant ancilla coupling provide promising avenues for implementing this minimal dissipative protocol without demanding complex multi-photon drives or continuous monitoring.
In superconducting circuits, phase rotations correspond to virtual frame updates with zero physical duration, and fast ancilla reset is well established~\cite{Geerlings2013}.

\begin{acknowledgments}
S.K. and K.P. acknowledge project CZ.02.01.01/00/22\_008/0004649 (QUEENTEC) of the EU and the Czech Ministry of Education, Youth and Sports (MEYS). R.F. acknowledges the European Union's HORIZON Research and Innovation Actions under Grant Agreement no.~101080173 (CLUSTEC), project 8C24003 of the MEYS CR, and the EU under Grant Agreements no.~101017733 and 731473 (CLUSSTAR). D.M. acknowledges project No.~LUC25006 of MEYS Czech Republic and project 9C26002 of the MEYS CR. A.R. acknowledges support from the NSRF via the Research and Innovation Acceleration Agency for Competitiveness and Area Development (RCAD) (Program Management Unit for Frontier Brainpower and Future Industries) [grant number B39G690076], and the Thailand Science Research and Innovation Fundamental Fund, fiscal year 2026 [grant number TUFF 49/2569]. The authors declare that large language models were used solely for text editorial suggestions, with all scientific calculations, derivations, and interpretations performed and verified directly by the authors.
\end{acknowledgments}

\bibliographystyle{unsrt}
\bibliography{references_v27}

\FloatBarrier

\appendix
\setcounter{secnumdepth}{1}

\section{Kraus map and Non-Gaussian Modulation Mechanism}
\label{append:kraus}

For resonant Jaynes-Cummings coupling in the interaction picture $\hat{H}_{\text{JC}} = g(\hat{a}\hat{\sigma}_+ + \hat{a}^\dagger\hat{\sigma}_-)$ acting for duration $t$ (dimensionless duration $\tau \equiv gt$), the unitary evolution operator is $\hat{U}(\tau) = \exp(-i\hat{H}_{\text{JC}}t)$. Given the initial joint state $\rho_{\text{tot}} = \rho_{\text{osc}} \otimes |g\rangle\langle g|$, tracing out the ground-state ancilla yields the completely positive trace-preserving reduced channel:
\begin{align}
  \rho'_{\text{osc}} &= \mathrm{Tr}_{\text{qb}}\!\left[\hat{U}(\tau)(\rho_{\text{osc}} \otimes |g\rangle\langle g|)\hat{U}^\dagger(\tau)\right] \nonumber \\
  &= \hat{M}_g(\tau) \rho_{\text{osc}} \hat{M}_g^\dagger(\tau) + \hat{M}_e(\tau) \rho_{\text{osc}} \hat{M}_e^\dagger(\tau).
\end{align}
We extract the Kraus operators $\hat{M}_s(\tau) = \langle s|\hat{U}(\tau)|g\rangle$ ($s \in \{g,e\}$) directly from the unitary evolution:
\begin{align}
  \hat{M}_g(\tau) &= \langle g| \hat{U}(\tau) |g\rangle = \sum_{n=0}^\infty \cos(\tau\sqrt{n})\,|n\rangle\langle n|, \label{eq:Mg_expansion} \\
  \hat{M}_e(\tau) &= \langle e| \hat{U}(\tau) |g\rangle = -i \sum_{n=1}^\infty \sin(\tau\sqrt{n})\,|n-1\rangle\langle n|. \label{eq:Me_expansion}
\end{align}
The action of this channel on the density matrix elements $\rho_{nm} = \langle n|\rho_{\text{osc}}|m\rangle$ in the Fock basis is given explicitly by:
\begin{align}
  \rho'_{nm} &= \cos(\tau\sqrt{n})\cos(\tau\sqrt{m})\,\rho_{nm} \nonumber \\
  &\quad + \sin(\tau\sqrt{n+1})\sin(\tau\sqrt{m+1})\,\rho_{n+1,m+1}.
  \label{eq:kraus_fock_elements}
\end{align}

The operator \mbox{$\hat{M}_g(\tau)$} acts as a diagonal phase-and-amplitude modulator on the Fock basis, whereas $\hat{M}_e(\tau)$ is a lowering operator that transfers photon excitations to the ancilla qubit. The joint evolution transfers intermediate Fock components $n$ to the ancilla excited state $|e\rangle$ with rates $\Omega_n = g\sqrt{n}$, where intermediate excitations are lowered via the non-unitary Kraus channel $\hat{M}_e$ upon tracing out the ancilla. Across successive cycles, the interleaved Gaussian unitaries $\hat{U}_{G,j}$ dynamically repump and realign residual amplitudes, driving destructive interference that concentrates coherence into the target components without requiring steady-state dark trapping.

To analyze the nonlinear dissipative process under decoherence, we use the Lindblad master equation:
\begin{equation}
  \frac{d\rho}{dt} = -i[\hat{H}_{\text{JC}}, \rho] + \mathcal{L}_{\text{osc}}\rho + \mathcal{L}_{\text{qb}}\rho.
  \label{eq:LindbladMaster}
\end{equation}
The oscillator dissipation accounts for single-photon loss at rate $\kappa$ and thermal noise with mean photon number $\bar{n}_{\text{th}}$:
\begin{equation}
  \mathcal{L}_{\text{osc}}\rho = \kappa(1+\bar{n}_{\text{th}})\mathcal{D}[\hat{a}]\rho + \kappa\bar{n}_{\text{th}}\mathcal{D}[\hat{a}^\dagger]\rho,
  \label{eq:osc_dissipator}
\end{equation}
where $\mathcal{D}[\hat{L}]\rho = \hat{L}\rho\hat{L}^\dagger - \frac{1}{2}\{\hat{L}^\dagger\hat{L}, \rho\}$. The ancillary qubit imperfections are modeled by energy relaxation ($T_1$) and pure dephasing ($T_\phi$):
\begin{equation}
  \mathcal{L}_{\text{qb}}\rho = \frac{1}{T_1}\mathcal{D}[\hat{\sigma}_-]\rho + \frac{1}{2T_\phi}\mathcal{D}[\hat{\sigma}_z]\rho.
\end{equation}
For the robustness analysis, the effective coherence time $T_2$ is defined via $1/T_2 = 1/(2T_1) + 1/T_\phi$.
Equation~(\ref{eq:LindbladMaster}) governs the joint system dynamics and environmental decoherence during the interaction intervals $t_j$.

\subsection*{Non-Equilibrium Multi-Step Mechanism and Incommensurate Spectrum}

For multi-component target superpositions, such as the binomial code state $|+_L\rangle$ $= \frac{1}{2}(|0\rangle+|4\rangle) + \frac{1}{\sqrt{2}}|2\rangle$, the system involves multiple Rabi frequencies $\Omega_n = g\sqrt{n}$, specifically $\Omega_0 = 0$, $\Omega_2 = g\sqrt{2}$, and $\Omega_4 = 2g$. Because the frequency ratio $\sqrt{2}/\sqrt{4} = 1/\sqrt{2}$ is irrational, the JC spectrum is incommensurate: no single interaction duration $\tau_j$ can simultaneously satisfy $\sin(\tau_j\sqrt{2}) \approx 0$ and $\sin(2\tau_j) \approx 0$ while maintaining non-zero excitation transfer $\sin(\tau_j\sqrt{n}) \neq 0$ for all unwanted intermediate Fock components $n \in \{1, 3\}$.

In continuous dissipative state engineering (DSE)~\cite{Poyatos1996PRL,Mirrahimi2014NJP,Leghtas2015Science}, stabilizing multi-component manifolds requires engineered reservoirs with high-order multi-photon jump operators, which inherently suffer from dark-state trapping, parasitic Kerr shifts, and slow asymptotic relaxation rates. Our protocol resolves this challenge through a non-equilibrium, multi-step sequence that combines non-unitary Kraus projection with active symplectic steering:
\begin{enumerate}[label=(\roman*)]
  \item \textbf{Kraus Modulation (\mbox{$\hat{M}_g, \hat{M}_e$}):} At each cycle $j$, the JC interaction for duration $\tau_j$ transfers intermediate Fock populations into the ancillary excited state $|e\rangle$ according to Eq.~(\ref{eq:kraus_fock_elements}). Tracing out the ancilla ($\mathrm{Tr}_{\mathrm{qb}}$) non-unitarily discards these correlated excitations into the environment, preventing information backflow.
  \item \textbf{Symplectic Phase Re-alignment ($\hat{U}_{G,j}$):} While $\hat{M}_e$ deposits a fraction of lowered population into $|n-1\rangle\langle n-1|$, the subsequent active Gaussian unitary $\hat{U}_{G,j} = \hat{R}(\theta_j)\hat{S}(\xi_j)\hat{D}(\beta_j)$ displaces, squeezes, and rotates the oscillator state. This realigns the relative phases of the residual amplitudes in phase space.
  \item \textbf{Destructive Interference without Trapping:} In the subsequent interaction step $j+1$, the realigned residual components interfere destructively with unwanted Fock amplitudes, transferring remaining errors into the non-unitary $|e\rangle$ channel. Because the operations are applied in discrete finite steps rather than continuously, the dynamics avoids micromaser-like trapping states~\cite{Filipowicz1986PRA} and converges to the target code within $k=5$ cycles.
\end{enumerate}

\subsection*{Algebraic Commutation and Intermediate Rotation Absorption}

Operationally, the intermediate phase rotation $\hat{R}(\theta_j) = \exp(-i\theta_j\hat{n})$ can be absorbed via phase-reparameterization of subsequent optimized displacements and squeezings, or deferred to the final step $j = k$, simplifying intermediate experimental control requirements without compromising fidelity. The complex parameter of displacement enables this directly through algebraic commutation identities:
\begin{align}
  \hat{R}(\theta)\hat{D}(\beta) &= \hat{D}(\beta e^{-i\theta})\hat{R}(\theta), \label{eq:rot_disp} \\
  \hat{R}(\theta)\hat{S}(\xi)\hat{R}^\dagger(\theta) &= \hat{S}(\xi e^{-2i\theta}), \label{eq:rot_sqz} \\
  \hat{R}(\theta)\mathcal{E}_{\text{JC}}(\rho)\hat{R}^\dagger(\theta) &= \mathcal{E}_{\text{JC}}(\hat{R}(\theta)\rho\hat{R}^\dagger(\theta)), \label{eq:rot_jc}
\end{align}
where Eq.~(\ref{eq:rot_jc}) holds because $[\hat{n}, \hat{M}_g] = 0$ and the phase $e^{i\theta}$ from $\hat{R}\hat{M}_e\hat{R}^\dagger = \hat{M}_e e^{i\theta}$ cancels in the Kraus product $\hat{M}_e \rho \hat{M}_e^\dagger$. Commuting $\hat{R}(\theta_j)$ forward past the JC channel $\mathcal{E}_{\text{JC}}$ acts on subsequent operations from the right via $\hat{D}(\beta_{j'})\hat{R}(\theta_j) = \hat{R}(\theta_j)\hat{D}(\beta_{j'} e^{i\theta_j})$ and $\hat{S}(\xi_{j'})\hat{R}(\theta_j) = \hat{R}(\theta_j)\hat{S}(\xi_{j'} e^{2i\theta_j})$, shifting subsequent drive arguments by $\beta_{j'} \to \beta_{j'} e^{i\theta_j}$ and $\xi_{j'} \to \xi_{j'} e^{2i\theta_j}$ for all $j' > j$. Consequently, an experimentalist does not need an active phase-rotation gate at each intermediate cycle; all intermediate rotations are absorbed into the complex drive arguments of subsequent displacements and squeezings, leaving at most a single cumulative phase rotation $\hat{R}(\Theta_{\text{total}})$ ($\Theta_{\text{total}} = \sum_{j=1}^k \theta_j$) at the final step $j = k$.

In numerical optimization, however, explicitly retaining $\hat{R}(\theta_j)$ per cycle is mathematically advantageous: the single-mode Gaussian symplectic group $\mathrm{Sp}(2, \mathbb{R})$ has three generators, whereas squeezing $\hat{S}(\xi)$ provides two parameters $(r, \phi)$ that degenerate at $r = 0$ ($\hat{S}(0) \equiv \hat{I}$ for all $\phi$). Explicit parameterization of $\hat{R}(\theta_j)$ avoids coordinate singularities during gradient descent when pure vanishing squeezing occurs (in Table~\ref{tab:parameters} a non-zero rotation is applied only at the final step, $\theta_5 = \pi/4$). In hardware such as circuit QED, phase rotations are virtual frame updates with zero duration and zero decoherence overhead. Commuting rotations backward through the sequence via Eq.~(\ref{eq:rot_jc}) and absorbing them into earlier Gaussian drives terminates on the vacuum \mbox{$\hat{R}(\Theta)|0\rangle = |0\rangle$}. Physical phase-rotation gates are therefore completely redundant in hardware and can be absorbed into drive phases. Explicitly, replacing $\beta_j\to\beta_j e^{-i\theta_5}$ and $\xi_j\to\xi_j e^{-2i\theta_5}$ for all $j=0,\dots,5$, including the initial $\beta_0$ and $\xi_0$ (for Table~\ref{tab:parameters}, $\beta_j e^{-i\pi/4}$ and $\xi_j e^{-i\pi/2}$), gives the identical output state with no rotation at any step.

\setcounter{figure}{0}
\renewcommand{\thefigure}{S\arabic{figure}}

\section{Optimization Parameters and Convergence Dynamics}
\label{append:params}

The deterministic protocol relies on a sequence of $k=5$ interactions, interspersed with optimized active Gaussian unitary corrections. The total system evolution is governed by the parameter set $\{\beta_0, \xi_0\}$ for the initial state preparation, and the set $\{\theta_j, \beta_j, \xi_j, \tau_j\}$ for each step $j$ ($1 \le j \le k$).

The initial state is prepared as $|\psi_{osc}(0)\rangle = \hat{S}(\xi_0)\hat{D}(\beta_0)|0\rangle$. In each step $j$, the oscillator undergoes a JC interaction for dimensionless duration $\tau_j = g t_j$, followed by a Gaussian correction $\hat{U}_{G,j} = \hat{R}(\theta_j)\hat{S}(\xi_j)\hat{D}(\beta_j)$. The displacement and squeezing parameters are defined as complex numbers $\beta_j = |\beta_j|e^{i\phi_{\beta_j}}$ and $\xi_j = r_j e^{i\phi_j}$.

Because the sequential protocol operates at shallow circuit depth ($k=5$) with only $25$ real optimization parameters across initial preparation and 5 cycles, gradients remain well-conditioned. The non-unitary Kraus channel confines the state evolution, ensuring fast and robust convergence (typically $<100$ gradient iterations) from random seeds without encountering local trapping.

Table~\ref{tab:parameters} lists the numerically optimized parameters used to generate the vacuum-Fock superposition ($|0\rangle + |4\rangle)/\sqrt{2}$ presented in the main text.

\begin{table}[t]
\centering
\caption{\textbf{Optimized Protocol Parameters.} Parameters for the generation of the state $(|0\rangle + |4\rangle)/\sqrt{2}$ using $k=5$ steps, reaching $\mathcal{F} = 0.9926$ with purity $0.9872$. The dimensionless interaction time is defined as $\tau_j = g t_j$.
The initial state is $\hat{S}(\xi_0)\hat{D}(\beta_0)\ket{0}$ with $\xi_0 = r_0 e^{i\phi_0}$,
$r_0 = 0.1017$, $\phi_0 = 0$, and $\beta_0 = 0.1965$. Each step applies the interaction for a time $\tau_j$ followed by $\hat{R}(\theta_j)\hat{S}(\xi_j)\hat{D}(\beta_j)$ with $\xi_j = r_j e^{i\phi_j}$; $\phi_j$ is omitted where $r_j = 0$. The final rotation $\theta_5 = \pi/4$ sets the relative phase between $|0\rangle$ and $|4\rangle$; without it the output is aligned with $(|0\rangle - |4\rangle)/\sqrt{2}$. Note: Parameters from an extended numerical search with one-dimensional scans of each $\tau_j$ over $\tau_j \le 80$; the solution is not restricted to the short-interaction regime ($\tau_1 \approx 59.9$, $\tau_3 \approx 68.9$), where $\tau_j \le 80$ corresponds to $t_j \le 637\text{ ns}$ for $g/2\pi = 20\text{ MHz}$, well below cavity coherence times.
}
\label{tab:parameters}\label{tab:G_04_results}
\begin{tabular}{c c cc c c}
\toprule
\multirow{2}{*}{Step}
& \multirow{2}{*}{$\tau_j$}
& \multicolumn{2}{c}{$S(\xi_j)$}
& \multirow{2}{*}{$D(\beta_j)$}
& \multirow{2}{*}{$R(\theta_j)$} \\
\cmidrule(lr){3-4}
& & $r_j$ & $\phi_j$ & & \\
\midrule
1 & 59.9030 & 0.0807 & 0      & $0.6954$  & 0 \\
2 & 2.7873  & 0.1636 & 0      & $0.4912$  & 0 \\
3 & 68.8749 & 0.0404 & 0      & $0.7888$  & 0 \\
4 & 17.6676 & 0.0482 & 0      & $1.1160$  & 0 \\
5 & 32.5453 & 0.0252 & 3.1416 & $-0.3768$ & $\pi/4$ \\
\bottomrule
\end{tabular}
\end{table}

\subsection{Convergence of $(|0\rangle+|N\rangle)/\sqrt{2}$ and $(|1\rangle+|N\rangle)/\sqrt{2}$ Superpositions}

\begin{figure}[h]
  \centering
  \includegraphics[width=\columnwidth]{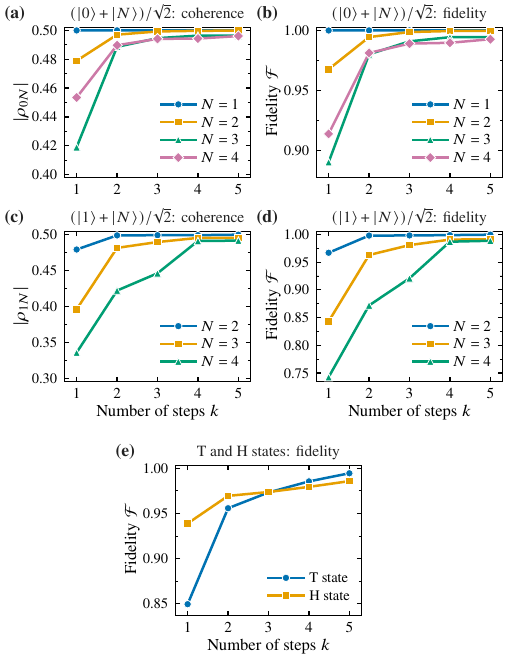}
  \caption{\textbf{Convergence of fidelity and coherence with the number of JC rounds $k$.} Highest fidelity found for each target with $k=1$--$5$ rounds, each followed by its Gaussian correction, with all rounds of a given $k$ optimized jointly; $k=5$ are the protocols used in the main text. (a) Coherence $|\rho_{0N}|$ and (b) fidelity of the states $(|0\rangle+|N\rangle)/\sqrt{2}$, $N=1$--$4$; (c) $|\rho_{1N}|$ and (d) fidelity of the states $(|1\rangle+|N\rangle)/\sqrt{2}$, $N=2$--$4$; (e) fidelity of the logical $T$- and $H$-states, including the free final phase-space rotation. Higher-order states start lower and need more rounds to reach the fidelity plateau, reflecting the increased dissipative work required: two rounds already give $\mathcal{F}>0.98$ for every $(|0\rangle+|N\rangle)/\sqrt{2}$ state, whereas the $(|1\rangle+|N\rangle)/\sqrt{2}$ states need two ($N=2$), three ($N=3$) and four ($N=4$) rounds, the $T$-state four, and the $H$-state five.}
  \label{fig:Evol}
\end{figure}

To characterize the efficiency of the sequential protocol, we analyze the scaling of state fidelity $\mathcal{F}$ and Fock-basis coherence $|\rho_{nm}|$ as a function of the total number of optimized JC rounds $k \in \{1, \dots, 5\}$ (with all rounds of a given $k$ optimized jointly). Figure~\ref{fig:Evol} illustrates this progression for the $|0\rangle+|N\rangle$, $|1\rangle+|N\rangle$, and logical magic states, respectively.

For both $|0\rangle+|N\rangle$ and $|1\rangle+|N\rangle$ superpositions, the protocol exhibits a rapid ascent in both fidelity and coherence. Low-order states ($N=1, 2$) reach their convergence plateau within $k=2$ rounds; remarkably, the $|0\rangle+|4\rangle$ state also already reaches $\mathcal{F} = 0.981$ at $k=2$ rounds (Fig.~\ref{fig:Evol}). In contrast, higher-order non-vacuum superpositions ($|1\rangle+|3\rangle, |1\rangle+|4\rangle$) as well as logical magic states require $k=3$ to $5$ rounds to surpass the $\mathcal{F} > 0.95$ threshold. %
This behavior stems from the increased number of destructive interference cycles required to suppress intermediate Fock populations and concentrate coherence into higher Fock-basis elements.

\section{Non-Vacuum Benchmarks: $(|1\rangle+|N\rangle)/\sqrt{2}$ Superpositions}
\label{appendix:appendix1N}

\begin{figure*}
  \centering
  \includegraphics[width=0.9\textwidth]{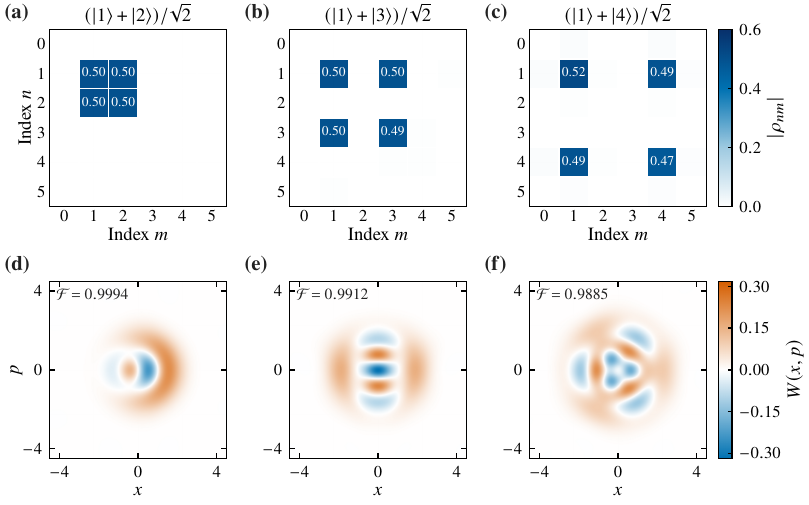}
  \caption{\textbf{Generated states $(|1\rangle+|N\rangle)/\sqrt{2}$ with their density matrices and Wigner functions ($k=5$ JC interactions).}
    \textbf{(a--f)} show the results of preparing these states for $N=2$, $3$, and $4$ (left to right; the target state is given above each column) using a deterministic approach based on five Jaynes--Cummings (JC) interactions.
    \textbf{(a), (b), and (c)} illustrate the absolute values of the density matrix elements $|\rho_{nm}|$ in the Fock basis for the generated states ($N=2,3,4$); only the block $0\le n,m\le 5$ is shown.
    \textbf{(d), (e), and (f)} display the Wigner functions $W(x,p)$ for each state, shown after the free phase rotation $e^{i\theta\hat{n}}$ that makes $\rho_{1N}$ real and positive.
  The fidelities with respect to the ideal states $(|1\rangle+|N\rangle)/\sqrt{2}$ are 0.9994 (purity $=0.9990$) for $N=2$, 0.9912 (purity $=0.9844$) for $N=3$, and 0.9885 (purity $=0.9797$) for $N=4$. These two-component states serve as elementary benchmarks away from the vacuum, whereas the binomial-code magic states in the main text combine three Fock components, $|0\rangle$, $|2\rangle$ and $|4\rangle$ (Fig.~\ref{fig:Magicstates}).}
  \label{fig:1Nstates}
\end{figure*}

To complement the method for general states other than the $|0\rangle+|N\rangle$ states, the protocol was applied to generate non-vacuum binary superpositions $(|1\rangle + |N\rangle)/\sqrt{2}$ with $N \in \{2, 3, 4\}$.
These two-component superpositions ($N \in \{2, 3, 4\}$) serve strictly as elementary non-vacuum benchmarks to test coherence synthesis away from the vacuum attractor. Within the fixed budget of $k=5$ JC cycles, standalone binary synthesis terminates at $N=4$, in contrast to the composite multi-component magic states in the main text where Fock components up to $n=5$ appear as residual out-of-subspace leakage.
Figure~\ref{fig:1Nstates} displays the density matrices and Wigner functions for $N=2,3,4$. The protocol achieves high fidelities: $\mathcal{F} = 0.9994, 0.9912, 0.9885$ for $N=2,3,4$.

\begin{figure}[h]
  \centering
  \includegraphics[width=\columnwidth]{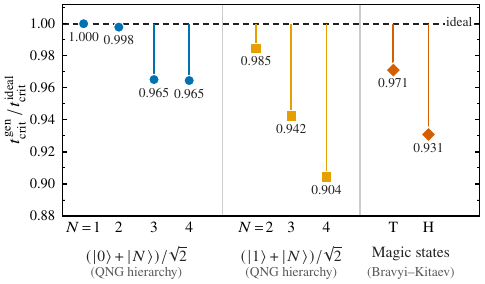}
  \caption{Thermal-depth ratio $t_{\rm crit}^{\rm gen}/t_{\rm crit}^{\rm ideal}$ of the generated states in a thermal bath with $\kappa=0.1$, $\bar n_{\rm th}=0.1$, compared to their ideal states: for the $(|0\rangle+|N\rangle)/\sqrt{2}$ and $(|1\rangle+|N\rangle)/\sqrt{2}$ states at the QNG hierarchy threshold on $|\rho_{mn}|$ defined in~\cite{kovalenko_quantum_2025}, and for the logical $T$- and $H$-states at the Bravyi--Kitaev threshold on the fidelity~\cite{BravyiPRA2005}, $\mathcal{F}\approx0.859$ for both. $t_{\rm crit}$ is the time at which the coherence (fidelity) first falls below the threshold; the dashed line marks the ideal state and each stem the shortfall. The ratio quantifies how closely the generated states approach the ideal resilience against thermal decoherence: all keep at least $90\%$ of the ideal lifetime, and the $(|0\rangle+|N\rangle)/\sqrt{2}$ states at least $95\%$.}
  \label{fig:thermal_ratio}
\end{figure}

Figures~\ref{fig:decoherence1N} and \ref{fig:thermal_ratio} illustrate the coherence decay of these $|1\rangle+|N\rangle$ states. Similar to the $|0\rangle+|N\rangle$ case, the $|1\rangle+|N\rangle$ states exhibit strong resilience against photon loss and thermal decoherence, with coherences $|\rho_{1N}|$ persisting above the QNG hierarchy thresholds for significant loss percentages.

\begin{figure*}
  \includegraphics[width=\textwidth]{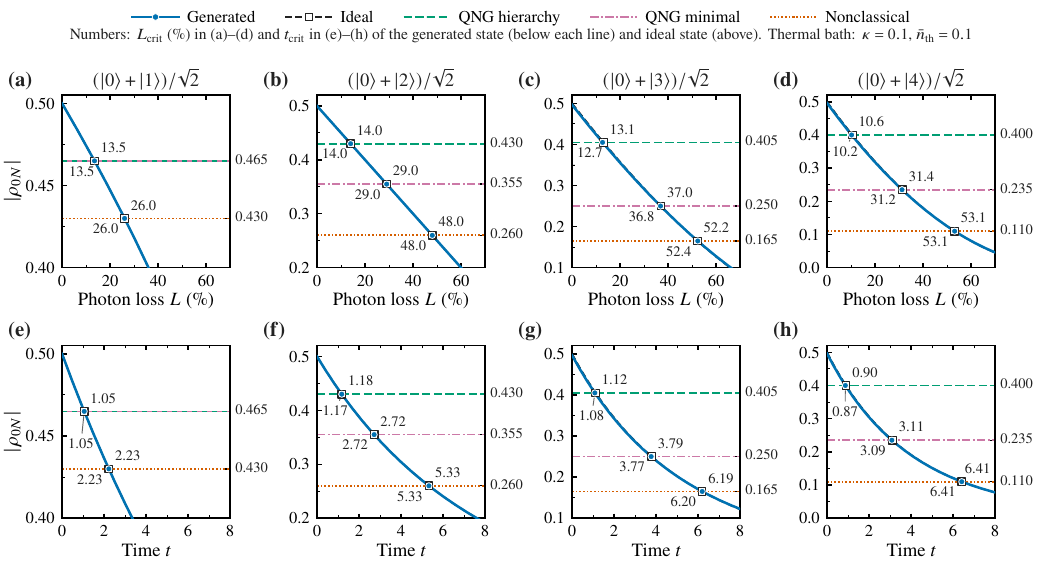}
    \caption{\textbf{Robustness of the generated states $(|0\rangle+|N\rangle)/\sqrt{2}$} ($N=1$--$4$, columns, with the target state given above each column) to photon loss (a--d) and
  thermal decoherence (e--h). $|\rho_{0N}|$ of the generated state (solid) and of the ideal target state (dashed) versus the photon-loss fraction $L=1-e^{-\kappa t}$ and versus time in a
  thermal bath with $\kappa=0.1$, $\bar n_{\rm th}=0.1$, both evolved with the oscillator dissipator of Eq.~(\ref{eq:osc_dissipator})
  (with
  $\bar n_{\rm th}=0$ for pure loss) acting on the prepared state. Horizontal lines mark the QNG hierarchy (dashed), QNG
  minimal (dash-dotted) and nonclassicality (dotted) thresholds, with their values on the right axis; for $N=1$ the two
  QNG thresholds coincide (alternating dashes). Circles (squares) mark the critical loss $L_{\rm crit}$ or time
  $t_{\rm crit}$ at which the generated (ideal) coherence falls below each threshold, with its value written below
  (above) the line. The generated coherence remains above the nonclassicality threshold up to
  $L_{\rm crit}=26.0\%$, $48.0\%$, $52.4\%$, $53.1\%$ and $t_{\rm crit}=2.23$, $5.33$, $6.20$, $6.41$ for $N=1$--$4$;
  for every threshold it stays within $0.5$ percentage points of the ideal $L_{\rm crit}$ and reaches at least
  $95\%$ of the ideal $t_{\rm crit}$.}
  \label{fig:decoherence0N}
\end{figure*}

\begin{figure*}
  \includegraphics[width=\textwidth]{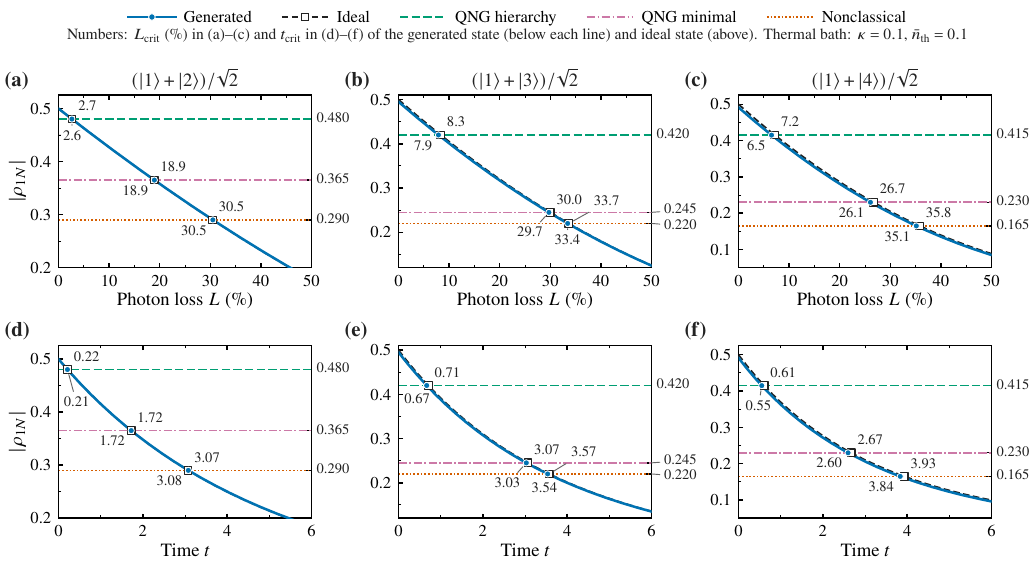}
  \caption{\textbf{Robustness of the generated states $(|1\rangle+|N\rangle)/\sqrt{2}$} ($N=2$--$4$, columns, with the target state given above each column) to photon loss (a--c) and
  thermal decoherence (d--f); noise models, lines and markers as in Fig.~\ref{fig:decoherence0N}, with
  $|\rho_{1N}|$ compared with the ideal target state. The generated coherence remains above
  the nonclassicality threshold up to $L_{\rm crit}=30.5\%$, $33.4\%$, $35.1\%$ and $t_{\rm crit}=3.08$, $3.54$,
  $3.84$ for $N=2$--$4$; for every threshold it stays within $0.7$ percentage points of the ideal $L_{\rm crit}$ and
  reaches at least $90\%$ of the ideal $t_{\rm crit}$.}
  \label{fig:decoherence1N}
\end{figure*}

Figures~\ref{fig:decoherence0N} and \ref{fig:decoherence1N} summarize these results, tracking the achieved coherence values against the theoretical nonclassical and QNG hierarchy bounds, while Figure~\ref{fig:coherence_thresholds} summarizes the initial coherence values prior to dissipation.
In all cases ($N=1$ to $4$), the generated states strictly surpass the highest-order QNG hierarchy threshold.
This confirms that the protocol successfully generates genuine quantum non-Gaussianity, clearly distinguishing the states from classical mixtures or lower-order non-Gaussian resources.

While the previous results assume ideal conditions, experimental realism requires accounting for dissipation. Comparing the thermal and damping depths of the $|1\rangle+|N\rangle$ superpositions to the $|0\rangle+|N\rangle$ states analyzed in Appendix~\ref{appendix:decay}, we observe that higher Fock components in $|1\rangle+|N\rangle$ states are intrinsically more fragile, leading to a sharper decline in the critical survival ratio $t_{\text{crit}}^{\text{gen}}/t_{\text{crit}}^{\text{ideal}}$ as $N$ increases. Despite these challenges, the generated $|1\rangle+|N\rangle$ states still successfully surpass genuine QNG thresholds under dissipation, proving that the protocol is not limited to vacuum-based superpositions.

\section{Coherence Decay of Prepared States}
\label{appendix:decay}\label{append:coherence}

\begin{figure}[h]
  \centering
  \includegraphics[width=\columnwidth]{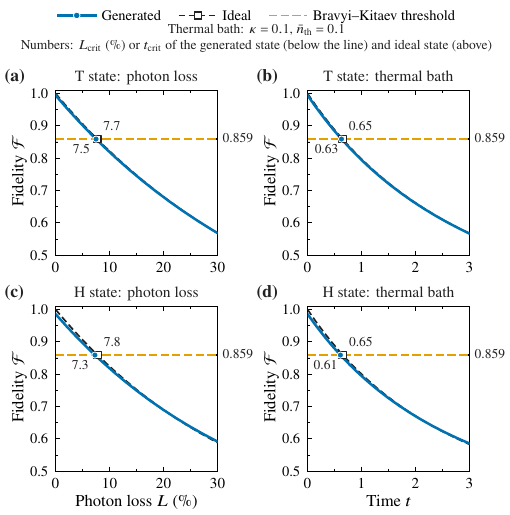}
  \caption{Fidelity of the generated T and H magic states (solid) and of the ideal states (dashed) under photon loss (a,c) and in a thermal bath with $\kappa=0.1$, $\bar n_{\rm th}=0.1$ (b,d), maximized over the free phase rotation. Dashed horizontal lines mark the Bravyi--Kitaev threshold $\mathcal{F}\approx0.859$~\cite{BravyiPRA2005}, the same for both states because they are related by a logical Clifford operation, with its value on the right axis; circles (squares) mark the critical loss $L_{\rm crit}$ or time $t_{\rm crit}$ at which the generated (ideal) fidelity falls below the threshold, with its value written below (above) the line. The generated states remain above threshold up to $L_{\rm crit}=7.5\%$ (T) and $7.3\%$ (H), compared with $7.7\%$ and $7.8\%$ for the ideal states, and up to  $t_{\rm crit}=0.629$ (T) and $0.609$ (H), compared with $0.648$ and $0.654$ for the ideal states.
}
  \label{fig:decoherence_TH}
\end{figure}

\subsection*{Initial Coherence vs. Thresholds}
Figure~\ref{fig:coherence_thresholds} summarizes the initial coherence values ($C_{0,N}$ and $C_{1,N}$) for all generated states immediately after the preparation sequence ($k=5$ steps), prior to any dissipation. The bar charts explicitly compare the achieved coherence against three critical thresholds: non-classicality, minimal quantum non-Gaussianity, and the specific $N$-order QNG hierarchy. In all cases ($N=1$ to $4$), the generated states surpass the highest relevant threshold (dashed line), confirming the successful deterministic generation of genuine multipartite-like quantum non-Gaussianity.

In this section, we present the comprehensive decoherence dynamics for the generated $|0\rangle+|N\rangle$ superpositions (Fig.~\ref{fig:decoherence0N}), logical magic states (Fig.~\ref{fig:decoherence_TH}), and $|1\rangle+|N\rangle$ superpositions (Fig.~\ref{fig:decoherence1N}), alongside the thermal-depth ratios (Fig.~\ref{fig:thermal_ratio}) quantifying resilience against environmental loss and thermal noise. For the $|0\rangle+|N\rangle$ states (Fig.~\ref{fig:decoherence0N}), coherence $|\rho_{0N}|$ remains above the classical bound for damping depths up to $26.0\%$ ($N=1$) and up to $53.1\%$ ($N=4$). Nonclassicality persists for interaction times up to $t_{\text{crit}} = 2.23$ ($N=1$) and $6.41$ ($N=4$) under thermal noise ($\bar{n}_{\text{th}}=0.1$), while genuine quantum non-Gaussianity at the highest hierarchy threshold persists up to $t_{\text{crit}} = 1.05$ ($N=1$) and $0.87$ ($N=4$), retaining over $95\%$ of the ideal state survival times.

\subsection*{Robustness of $|1\rangle+|N\rangle$ Superpositions}
Similar to the $|0\rangle+|N\rangle$ states and the logical magic states (Fig.~\ref{fig:decoherence_TH}), the $|1\rangle+|N\rangle$ superpositions exhibit significant resilience against environmental noise. Figure~\ref{fig:decoherence1N} illustrates the decay of the off-diagonal coherence element $|\rho_{1N}|$ under single-photon loss. While higher-$N$ states naturally decay faster due to their larger size, the generated states maintain coherence above the classical bound (dotted) and the minimal QNG bound (dash-dotted) for substantial loss percentages (up to $\approx 35.1\%$ for $N=4$).

Figure~\ref{fig:thermal_ratio} displays the ``thermal depth'' analysis for these states, tracking the evolution of coherence under continuous thermal noise ($\bar{n}_{\text{th}} = 0.1$). The ratio $t_{\text{crit}}^{\text{gen}} / t_{\text{crit}}^{\text{ideal}}$ remains high across all $N$ (exceeding $90\%$), indicating that the deterministically prepared states are not significantly more fragile than ideal theoretical states in a thermal environment. This confirms that the protocol distributes non-linearity effectively without introducing excessive susceptibility to decoherence.

\section{Coherence Flow and Correlation Erasure Analysis}
\label{append:coherence_flow}

To elucidate the physical mechanism of our protocol, specifically the roles of the Gaussian operations and the dissipative qubit reset, we analyze the evolution of the system using the Relative Entropy of Coherence $C_{\text{rel}}(\hat{\rho})$ \cite{baumgratz2014quantifying}. This measure is defined as:
\begin{equation}
  C_{\text{rel}}(\hat{\rho}) = S(\hat{\rho}_{\text{diag}}) - S(\hat{\rho}),
\end{equation}
where $S(\hat{\rho}) = -\text{Tr}[\hat{\rho}\log_2\hat{\rho}]$ is the von Neumann entropy in bits, and $\hat{\rho}_{\text{diag}}$ denotes the density matrix with all off-diagonal elements removed (in the Fock basis).

For the composite oscillator-qubit system, the total coherence can be decomposed into local contributions and mutual correlations:
\begin{equation}
  C_{\text{total}} = C_{\text{osc}} + C_{\text{qubit}} + C_{\text{corr}},
\end{equation}
where $C_{\text{osc}}$ and $C_{\text{qubit}}$ are the local coherences, and $C_{\text{corr}} = C_{\text{total}} - C_{\text{osc}} - C_{\text{qubit}}$ quantifies the coherence contained in inter-subsystem correlations.

\subsection{Correlation Erasure via Ancilla Reset and Purity Preservation}

While the non-Gaussian phase modulation is imparted by the action of $\hat{M}_g = \cos(gt_j\sqrt{\hat{n}})$, the transient JC interaction builds temporary correlations between the oscillator and ancilla before the reset. As quantified by $C_{\mathrm{loss}} = C_{\mathrm{qubit}} + C_{\mathrm{corr}}$ in Table~\ref{tab:coherence_values}, tracing out the ancilla non-unitarily resets inter-subsystem correlations without information backflow. While intermediate excitations transfer population into $|e\rangle$ via $\hat{M}_e$, at the final step ($j=5$) the joint state before reset factorizes: $|\Psi_{\mathrm{joint}}\rangle \approx |\psi_{\mathrm{osc}}\rangle \otimes (\sqrt{p_g}|g\rangle - i\sqrt{p_e}|e\rangle)$ due to the high collinearity of the normalized conditional states $|\tilde{\psi}_g\rangle \equiv \hat{M}_g|\psi\rangle/\sqrt{p_g}$ and $|\tilde{\psi}_e\rangle \equiv \hat{M}_e|\psi\rangle/\sqrt{p_e}$ ($|\langle\tilde{\psi}_g|\tilde{\psi}_e\rangle| \approx 0.990$, with measurement probabilities $p_s = \langle\psi|\hat{M}_s^\dagger\hat{M}_s|\psi\rangle$). Consequently, inter-subsystem correlation remains minimal ($C_{\mathrm{corr}} = 0.081\text{ bits} \approx 0$), preserving overall oscillator purity ($\mathrm{Tr}[\rho_{\mathrm{osc}}^2] \approx 0.987$) immediately upon ancilla reset while depositing $C_{\mathrm{qubit}} = 0.943\text{ bits}$ of coherence into the discarded qubit channel.

Our analysis reveals that the optimized Gaussian operations do not necessarily maximize the global coherence of the oscillator ($C_{\text{osc}}$). Instead, as quantified in Table~\ref{tab:coherence_values}, they act to \textit{redistribute} the coherence. The Gaussian steps reshape the state, suppressing irrelevant off-diagonal terms and concentrating coherence into the specific Fock superposition elements (e.g., $\rho_{04}$) required for the target state.

\begin{table}[t]
\centering
\caption{\textbf{Evolution of Relative Entropy of Coherence.} Relative entropy of coherence $C(\rho) = S(\Delta[\rho]) - S(\rho)$, in bits and in the Fock ($\otimes$ qubit) basis, along the generation of $(|0\rangle + |4\rangle)/\sqrt{2}$ with the parameters of Table~\ref{tab:parameters}. After the interaction of step $j$, with dimensionless interaction time $\tau_j = g t_j$, $C_{\text{total}}$ is the joint oscillator--qubit coherence, $C_{\text{osc}}$ and $C_{\text{qubit}}$ are those of the reduced states, $C_{\text{corr}} = C_{\text{total}} - C_{\text{osc}} - C_{\text{qubit}}$ is the correlated coherence, and $C_{\text{loss}} = C_{\text{total}} - C_{\text{osc}}$ is the coherence discarded when the qubit is traced out. $C_{\text{osc}}^{\text{final}}$ is the oscillator coherence after the Gaussian operation (GO) of step $j$.
}
\label{tab:coherence_values}
\begin{tabular}{c c ccccc c}
\toprule
\multirow{2}{*}{Step}
& \multirow{2}{*}{$\tau_j$}
& \multicolumn{5}{c}{After interaction}
& After GO \\
\cmidrule(lr){3-7} \cmidrule(lr){8-8}
& & $C_{\text{total}}$ & $C_{\text{osc}}$ & $C_{\text{qubit}}$ & $C_{\text{corr}}$ & $C_{\text{loss}}$ & $C_{\text{osc}}^{\text{final}}$ \\
\midrule
Initial & -- & 0.253 & 0.253 & 0.000 & 0.000 & -- & -- \\
1       & 59.9030 & 0.263 & 0.242 & 0.015 & 0.005 & 0.020 & 0.805 \\
2       & 2.7873 & 0.920 & 0.728 & 0.190 & 0.002 & 0.192 & 0.577 \\
3       & 68.8749 & 0.583 & 0.563 & 0.006 & 0.015 & 0.021 & 1.782 \\
4       & 17.6676 & 2.011 & 1.289 & 0.701 & 0.021 & 0.722 & 2.701 \\
5       & 32.5453 & 3.234 & 2.211 & 0.943 & 0.081 & 1.023 & 1.005 \\

\bottomrule
\end{tabular}
\end{table}

\section{Phase sensitivity of the generated states}
\label{append:Fisher}

To further quantify the phase-sensitivity utility of the states generated by our protocol, we evaluate their phase sensitivity using both the quantum Fisher information (QFI) and the classical Fisher information (CFI). The QFI provides an ultimate upper bound on precision, while the CFI demonstrates the practically achievable sensitivity using an optimal projective measurement.

\begin{figure*}
  \includegraphics[width=\textwidth]{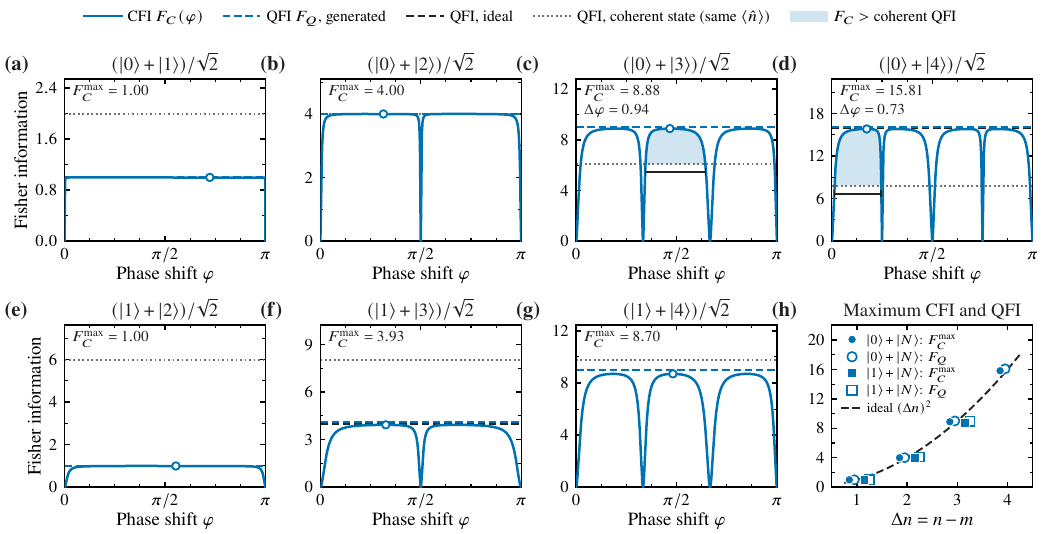}
  \caption{\textbf{Phase sensitivity of the generated states $(|0\rangle+|N\rangle)/\sqrt{2}$ and $(|1\rangle+|N\rangle)/\sqrt{2}$.} Phase-dependent classical Fisher information for the generated states $(|0\rangle+|N\rangle)/\sqrt{2}$, (a)--(d) for $N=1$--$4$, and $(|1\rangle+|N\rangle)/\sqrt{2}$, (e)--(g) for $N=2$--$4$, with the target state given above each panel. The solid blue curves show the classical Fisher information $F_C(\varphi)$ obtained from the generated mixed states after phase encoding by $\hat{n}$, using a projective measurement onto the corresponding ideal superposition. The blue dashed lines denote the mixed-state quantum Fisher information of the generated states, while the black dashed lines indicate the quantum Fisher information $(n-m)^2$ of the corresponding ideal superpositions. The gray dotted lines show the coherent-state benchmark with the same mean photon number; the shaded region marks the phase window of width $\Delta\varphi$ around the CFI maximum where $F_C$ exceeds it. For $N=2$ the maximum CFI exceeds the coherent-state value by only $10^{-4}$, so no window is marked. For the ideal states, the chosen projective measurement saturates the quantum Fisher information at the optimal phase points. The maximum CFI ratios relative to the ideal QFI are $F_C^{\max}/F_Q^{\mathrm{ideal}} = 0.9999$, 0.9994, 0.9869, and 0.9882 for $(|0\rangle+|N\rangle)/\sqrt{2}$ with $N=1$, $2$, $3$, and $4$, and 0.9984, 0.9836, and 0.9666 for $(|1\rangle+|N\rangle)/\sqrt{2}$ with $N=2$, $3$, and $4$, respectively. (h) Maximum CFI (filled) and mixed-state QFI (open) of the states $(|0\rangle+|N\rangle)/\sqrt{2}$ (circles) and $(|1\rangle+|N\rangle)/\sqrt{2}$ (squares) versus the photon-number difference $\Delta n = n-m$, with the ideal $(\Delta n)^2$. Each state is shown after the free phase rotation that makes $\rho_{mn}$ real and positive, which only shifts the curves in $\varphi$.}
  \label{fig:fisher}
\end{figure*}

In Fig.~\ref{fig:fisher}, we present the phase-dependent CFI and mixed-state QFI for the generated $|0\rangle+|N\rangle$ and $|1\rangle+|N\rangle$ superposition states. The results show that the CFI closely approaches the ideal QFI at optimal phase choices ($F_C^{\max}/F_Q^{\mathrm{ideal}} = 0.9999, 0.9994, 0.9869, 0.9882$ for the $|0\rangle+|N\rangle$ states ($N=1\text{--}4$), and $0.9984, 0.9836, 0.9666$ for the $|1\rangle+|N\rangle$ states ($N=2\text{--}4$)), indicating that phase information can be efficiently extracted. For states with residual population in higher Fock components ($n > N$), the usable sensitivity is rigorously bounded by the classical Fisher information.

Figure~\ref{fig:fisher}(h) summarizes these findings, displaying the maximum CFI and corresponding QFI as a function of the photon-number difference $\Delta n = n-m$ for both families. The steady scaling confirms that the deterministically generated high-Fock superpositions are robust resources for phase estimation.

\onecolumngrid

\end{document}